\documentclass[aps,amsmath,amssymb,twocolumn]{revtex4-1}

\usepackage{graphicx}
\usepackage{dcolumn}
\usepackage{bm}
\usepackage[normalem]{ulem}
\usepackage{color}
\usepackage{amsmath}

\begin{document}

\title{Tunneling processes between Yu-Shiba-Rusinov bound states}

\author{A. Villas$^{1,\dagger}$}

\author{R.~L. Klees$^{2,\dagger}$}

\author{G. Morr\'as$^{1}$}

\author{H. Huang$^{3}$}

\author{C.~R. Ast$^{3}$}

\author{G. Rastelli$^{2,4,5}$}

\author{W. Belzig$^{2}$}

\author{J.~C. Cuevas$^{1}$}

\affiliation{$^1$Departamento de F\'{\i}sica Te\'orica de la Materia Condensada
and Condensed Matter Physics Center (IFIMAC), Universidad Aut\'onoma de Madrid,
E-28049 Madrid, Spain}

\affiliation{$^2$Fachbereich Physik, Universit{\"a}t Konstanz, D-78457 Konstanz, Germany}

\affiliation{$^3$Max-Planck-Institut f\"ur Festk\"orperforschung, Heisenbergstra{\ss}e 1, 
70569 Stuttgart, Germany}

\affiliation{$^4$Zukunftskolleg, Universit{\"a}t Konstanz, D-78457 Konstanz, Germany}

\affiliation{$^5$INO-CNR BEC Center and Dipartimento di Fisica, Universit\`{a} di Trento, I-38123 Povo, Italy}

\date{\today}

\begin{abstract}
Very recent experiments have reported the tunneling between Yu-Shiba-Rusinov (YSR) bound states at the atomic scale. These 
experiments have been realized with the help of a scanning tunneling microscope where a superconducting tip is functionalized 
with a magnetic impurity and is used to probe another magnetic impurity deposited on a superconducting substrate. In this way 
it has become possible to study for the first time the spin-dependent transport between individual superconducting bound states. 
Motivated by these experiments, we present here a comprehensive theoretical study of the tunneling processes between YSR bound 
states in a system in which two magnetic impurities are coupled to superconducting leads. Our theory is based on a combination 
of an Anderson model with broken spin degeneracy to describe the impurities and nonequilibrium Green's function techniques to 
compute the current-voltage characteristics. This combination allows us to describe the spin-dependent transport for an arbitrary 
strength of the tunnel coupling between the impurities. We first focus on the tunnel regime and show that our theory naturally 
explains the experimental observations of the appearance of current peaks in the subgap region due to both the direct and thermal 
tunneling between the YSR states in both impurities. Then, we study in detail the case of junctions with increasing transparency, 
which has not been experimentally explored yet, and predict the occurrence of a large variety of (multiple) Andreev reflections 
mediated by YSR states that give rise to a very rich structure in the subgap current. In particular, we predict the occurrence 
of multiple Andreev reflections that involve YSR states in different impurities. These processes have no analogue in single-impurity 
junctions and they are manifested as current peaks with negative differential conductance for subgap voltages. Overall, our work 
illustrates the unique physics that emerges when the spin degree of freedom is added to a system with superconducting bound states.
\end{abstract}

\maketitle

\section{Introduction}

In recent years, the competition between magnetism and superconductivity has been extensively studied at the atomic scale with 
the help of the scanning tunneling microscope (STM). With this instrument it is possible to manipulate individual magnetic atoms 
and molecules and study the electronic transport through them when they are deposited on a superconducting substrate. In these 
single-impurity systems, the combination of spin-dependent scattering and superconductivity leads to the appearance of the so-called 
Yu-Shiba-Rusinov (YSR) states \cite{Yu1965,Shiba1968,Rusinov1969}, which are superconducting bound states with unique properties 
such as their spin polarization. Many STM-based experiments have demonstrated the existence of these bound states and, in turn, 
have elucidated many of their basic properties \cite{Yazdani1997,Ji2008,Franke2011,Menard2015,Ruby2015,Hatter2015,Ruby2016,
Randeria2016,Choi2017,Cornils2017,Hatter2017,Farinacci2018,Brand2018,Malavolti2018,Kezilebieke2019,Senkpiel2019,Schneider2019,
Liebhaber2020,Huang2020b,Odobesko2020}, for a recent review see Ref.~\cite{Heinrich2018}. Part of the interest in the physics of YSR 
states lies in the fact that they can be viewed as building blocks to create Majorana states in designer structures such as chains of 
magnetic impurities \cite{Nadj-Perge2014,Ruby2015b,Kezilebieke2018,Ruby2017,Ruby2018}.

Very recently, it has been experimentally demonstrated that a superconducting STM tip can be decorated with a magnetic impurity 
that then features YSR states \cite{Huang2020a}. More importantly, this YSR-STM can, in turn, be used to probe other magnetic 
impurities deposited on a superconducting substrate and that also features YSR states. In this way, the experiments realized for the 
first the time the tunneling between individual superconducting bound states at the atomic scale, which is the ultimate limit for 
quantum transport. Additionally, it has been shown that the YSR-STM can be used to measure the intrinsic lifetime of YSR states 
and that the tunnel current exhibits peaks in the subgap region due to direct and thermal tunneling between the YSR in both impurities
\cite{Huang2020a}. In particular, those current peaks can be used to extract information about the relative orientation between 
the impurity spins \cite{Huang2020c}. In fact, this system represents an ideal platform to explore the interplay between spin-dependent 
transport and superconductivity, which lies at the heart of the field of superconducting spintronics \cite{Linder2015,Eschrig2015,
Holmqvist2018}. On the other hand, it is obvious that the YSR-STM may have important implications for spin-polarized scanning tunneling 
microscopy and the study of atomic-scale magnetic structures, as it has been nicely demonstrated in Ref.~\cite{Schneider2020}.

Another exciting possibility that the YSR-STM opens up is the study of the interplay between superconducting bound states and 
(multiple) Andreev reflections in a situation never explored before and in which the spin degree of freedom plays a central role. 
Let us recall that in a junction with at least one superconducting electrode, an Andreev reflection consists of a tunneling process 
in which an electron coming from a normal metal is reflected as a hole of opposite spin transferring a Cooper pair into the superconductor. 
In the absence of in-gap bound states, this process dominates the subgap transport. If the junction features two superconducting leads, 
one can additionally have multiple Andreev reflections (MARs) in which quasiparticles undergo a cascade of Andreev reflections that 
give rise to a very rich subgap structure in the current-voltage characteristics. The microscopic theory of MARs for spin-degenerate 
quantum point contacts was developed in the mid-1990s \cite{Averin1995,Cuevas1996}, and it was first quantitatively confirmed in the 
context of superconducting atomic-size contacts with the help of break-junction techniques and the STM \cite{Scheer1997,Scheer1998}. 
In recent years, different STM experiments in the context of magnetic impurities on superconducting surfaces and using superconducting 
tips have revealed signatures of the interplay between YSR bound states and Andreev reflections \cite{Ruby2015,Randeria2016,Farinacci2018,
Brand2018,Huang2020c}. From the theory side, we have recently put forward a model to describe this interplay in single-impurity junctions 
and have shown how the spin degree of freedom leads to MAR processes that have no analogue in nonmagnetic systems. The qualitative predictions 
of this theory have been experimentally confirmed \cite{Huang2020c}. The goal of this work is to extend that theoretical analysis to 
the two-impurity case in order to elucidate the different tunneling processes that can take place between YSR states.

\begin{figure}[t]
\includegraphics[width=0.95\columnwidth,clip]{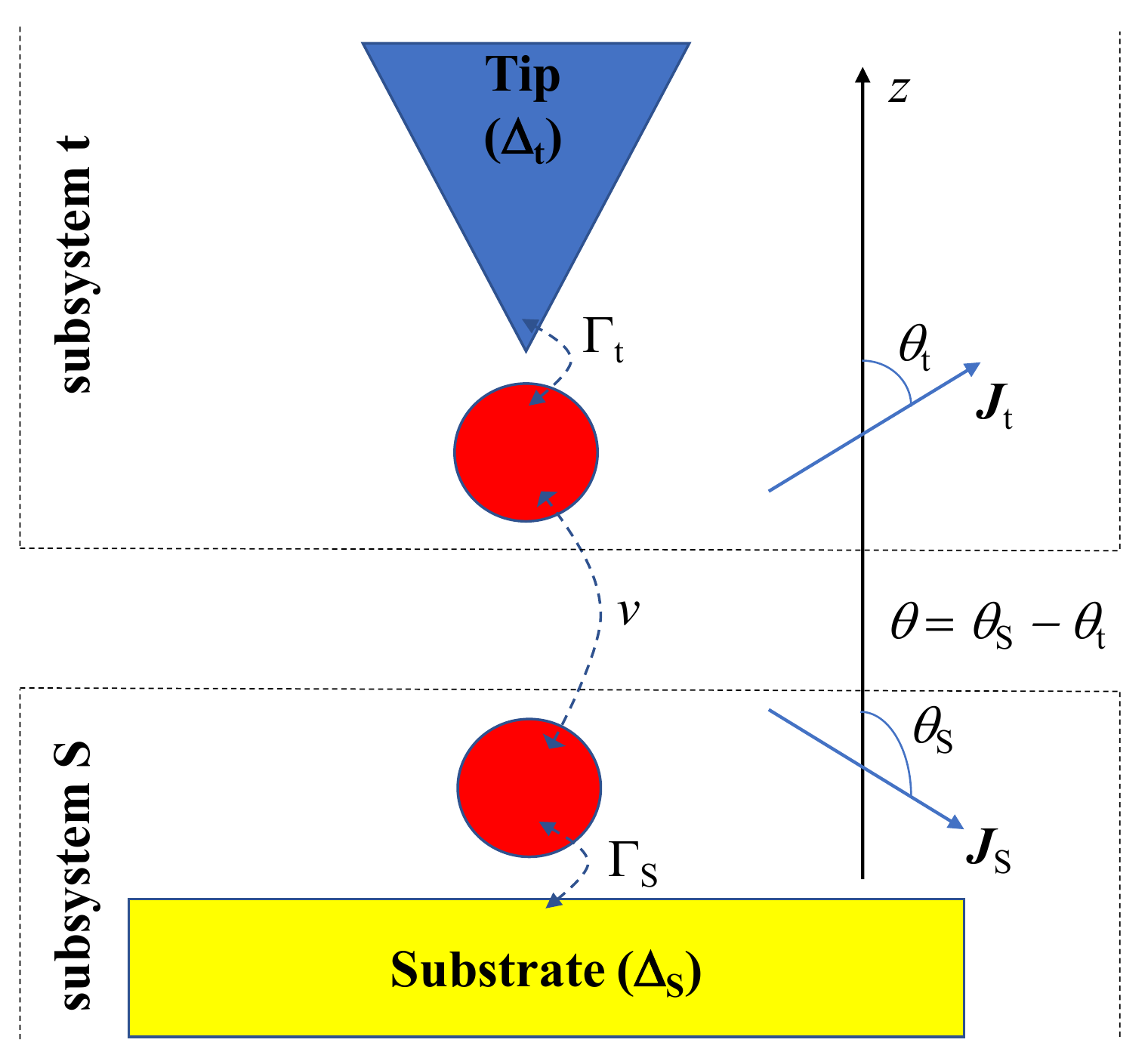}
\caption{Schematic representation of the system under study. Two magnetic impurities are respectively coupled to a superconducting 
substrate and to an STM tip that is also superconducting. The tunneling rates $\Gamma_{\rm t}$ and $\Gamma_{\rm S}$ measure the 
strength of the coupling of the impurity to the tip and substrate, respectively, $\Delta_{\rm t}$ and $\Delta_{\rm S}$ are 
the corresponding superconducting gaps, and $v$ is the hopping matrix element describing the tunnel coupling between the impurities.
These impurities have magnetizations $\boldsymbol{J}_\mathrm{t}$ and $\boldsymbol{J}_\mathrm{S}$ forming angles $\theta_\mathrm{t}$ 
and $\theta_\mathrm{S}$ with the quantization axis ($z$-axis) and their relative orientation is denoted by $\theta = 
\theta_\mathrm{S} - \theta_\mathrm{t}$.}
\label{fig-system}
\end{figure}

In this work we present a systematic study of the tunneling processes between YSR bound states in a system comprising two magnetic 
impurities that are coupled to their respective superconducting electrodes, see Fig.~\ref{fig-system}. Our theory is based on the use 
of a mean-field Anderson model with broken spin symmetry to describe the magnetic impurities and we employ the Keldysh formalism to 
compute the current-voltage characteristics for arbitrary junction transmission, i.e., to any order in the tunnel coupling between 
the two impurities. To illustrate the power of our model, we first focus on the analysis of the tunnel regime in which the charge 
transport is completely dominated by tunneling of single quasiparticles. In this regime, we naturally explain the basic observations 
reported in Refs.~\cite{Huang2020a,Huang2020c} concerning the presence of current peaks with huge negative differential conductance 
in the gap region. As explained in Refs.~\cite{Huang2020a,Huang2020c}, those peaks can be attributed to the direct and thermal tunneling 
between the YSR states in both impurities and their heights contain sufficient information to extract the relative orientation of the 
impurity spins. More importantly, we also study in detail how the transport characteristics change upon increasing the junction 
transparency and predict the occurrence of several families of MARs that give rise to an extremely rich subgap structure in the current 
and differential conductance. In particular, we find a series of MARs that start and end in YSR bound states, which are not possible in 
the case of single-impurity junctions. The signature of these YSR-mediated MARs is a series of current peaks at certain subgap voltages 
determined by the energy of the YSR states in both impurities. All the predictions put forward in this work can, in principle, be 
verified with the exact system investigated in Refs.~\cite{Huang2020a,Huang2020c}. 

The rest of the manuscript is organized as follows. In Sec.~\ref{sec-theory} we describe the system under study and present the 
model and theoretical tools that we have employed to study the electronic transport in our two-impurity superconducting system.
In Sec.~\ref{sec-tunnel-regime} we focus on the tunnel regime and show how our theory nicely explains all the basic observations 
reported in Refs.~\cite{Huang2020a,Huang2020c}. Then, in Sec.~\ref{sec-MARs} we present a detailed study of the subgap transport 
in junctions with a moderate-to-high transmission and analyze the interplay between MARs and YSR states. Finally, in 
Sec.~\ref{sec-conclusions} we summarize our main conclusions.

\section{System under study and theoretical approach} \label{sec-theory}

The goal of this work is to elucidate the different tunneling processes that can occur between YSR states. As explained in the 
introduction, these bound states appear in single magnetic impurities (atoms or molecules) coupled to superconducting electrodes 
and the tunneling between them is possible via direct inter-impurity coupling. This system has been realized with the 
help of an STM and, in this case, an impurity is coupled to the superconducting STM tip, while the other one is coupled to a 
superconducting substrate \cite{Huang2020a}, as we show schematically in Fig.~\ref{fig-system}. Thus, our technical goal is to 
compute the current-voltage characteristics in such a system and this section is devoted to a detailed description of the model 
and theoretical tools employed for this purpose. 

We consider the total system shown in Fig.~\ref{fig-system} and assume that the magnetic moments of the impurities form a relative 
angle $\theta$, which will be treated as a parameter of the model. Motivated by the experiments of Ref.~\cite{Huang2020a}, we shall 
assume that the impurities are strongly coupled to their respective electrode (STM tip and substrate), which is the regime in which 
the YSR states appear. In this sense, in order to describe the electronic transport in this system, it is natural to divide it into 
two subsystems, tip (t) and substrate (S), each one containing a magnetic impurity which is strongly coupled to a superconducting 
electrode. Moreover, we shall assume that the voltage drops at the interface between the two impurities. Such a system can be modeled 
by a generic point-contact Hamiltonian of the form
\begin{equation}
\label{eq-HLR}
H = H_\mathrm{t} + H_\mathrm{S} + V ,
\end{equation}
where $H_j$ with $j \in \{ \mathrm{t,S} \}$ describes the corresponding subsystem (i.e., an impurity coupled its superconducting 
electrode) and $V$ describes the tunnel coupling between these two subsystems. These different parts of the total Hamiltonian will 
be specified in the following subsections. 

\subsection{Bare Green's function of a magnetic impurity coupled to a superconductor and YSR states}

The impurities are described with a mean-field Anderson model with broken spin symmetry that was recently used to describe the 
role of the impurity-substrate coupling \cite{Huang2020b} and to elucidate the MARs that can take place in the electronic transport 
through a single magnetic impurity coupled to superconducting leads \cite{Villas2020}. This model has also
been successfully employed in the past to describe the observation of Andreev bound states in quantum dots coupled to superconducting 
leads and it has been shown to reproduce many of the salient features of the superconducting bound states predicted by more sophisticated 
many-body approaches \cite{Martin-Rodero2011,Martin-Rodero2012}. Within this model, we couple the magnetic 
impurity featuring a single energy level $U_j$ and a magnetization $\boldsymbol{J}_j = J_j ( \cos \theta_j \boldsymbol{e}_z + 
\sin \theta_j \boldsymbol{e}_x)$, where $\theta_j$ is the angle between the magnetization and a global quantization axis along 
the $z$ direction, to an s-wave superconductor. It is convenient to first focus on the individual subsystems described by $H_j$ and 
define the Hamiltonians and the effective Green's functions in each individual diagonal basis pointing along the direction of 
$\boldsymbol{J}_j$. The two separate bases are then simply related to the global quantization $z$ axis by a rotation of the above 
defined angle $\theta_j$ about the $y$ axis in spin space.

First, we define the spinors along the global quantization $z$ axis as
\begin{subequations}
	\label{eq-basisTot}
	\begin{align}
	\label{eq-basis}
	\tilde{\boldsymbol{d}}_{j}^{\dag} &= (d_{j\uparrow}^{ \dag}, d_{j\downarrow}^{\phantom \dag}, 
	d_{j\downarrow}^{ \dag}, - d_{j\uparrow}^{ \phantom\dag}) ,
	\\
	\tilde{\boldsymbol{c}}_{\boldsymbol{k}j}^\dag 
	&= 
	( c_{\boldsymbol{k}j\uparrow}^\dag , 
	c_{-\boldsymbol{k}j\downarrow}^{\phantom \dag} , 
	c_{\boldsymbol{k}j\downarrow}^\dag , 
	- c_{-\boldsymbol{k}j\uparrow}^{\phantom \dag} ) ,
	\end{align} 
\end{subequations}
which consist of annihilation (creation) operators $d_{j\sigma}^{(\dag)}$ and $c_{\boldsymbol{k}j\sigma}^{(\dag)}$ for electrons 
on the dot and the superconductor, respectively, with spin $\sigma \in \{ \uparrow , \downarrow\}$ and quasi-momentum $\boldsymbol{k}$. 
The Hamiltonian of the subsystem $j$ reads
\begin{align}
	H_j = H_{\mathrm{imp},j} + H_{\mathrm{elec},j} + V_j ,
\end{align}
where $H_{\mathrm{imp},j}$ describes the magnetic impurity in subsystem $j$, $H_{\mathrm{elec},j}$ describes the superconducting 
electrode in subsystem $j$, and $V_j$ describes their coupling in subsystem $j$. As it has been shown in Ref.~\cite{Villas2020}, 
by using the spinors in Eq.~\eqref{eq-basisTot} these Hamiltonians can be cast into the form
\begin{subequations}
	\label{eq-idividualHamilSubsys}
	\begin{align}
	H_{\mathrm{imp},j} &= \frac{1}{2} \tilde{\boldsymbol{d}}_j^\dag  \tilde{H}_{\mathrm{imp},j} 
	\tilde{\boldsymbol{d}}_j^{\phantom \dag} ,
	\\
	H_{\mathrm{elec},j} &= \frac{1}{2} \sum_{{\boldsymbol k}} \tilde{\boldsymbol{c}}_{\boldsymbol{k}j}^\dag 
	\tilde{H}_{\mathrm{elec}, \boldsymbol{k}j} \tilde{\boldsymbol{c}}_{\boldsymbol{k}j}^{\phantom \dag} ,
	\\
	V_j &= \frac{1}{2} \sum_{{\boldsymbol k}} \tilde{\boldsymbol{c}}_{\boldsymbol{k}j}^\dag \tilde{V}_{j} 
	\tilde{\boldsymbol{d}}_{j}^{\phantom \dag} + \frac{1}{2} \sum_{{\boldsymbol k}} \tilde{\boldsymbol{d}}_{j}^{ \dag}  
	\tilde{V}_{j}^\dag \tilde{\boldsymbol{c}}_{\boldsymbol{k}j}^{\phantom \dag} ,
	\end{align}
\end{subequations}
with the $4 \times 4$ matrix Hamiltonians
\begin{subequations}
	\label{eq-4by4Hamil}
	\begin{align}
		\label{eq-4by4HamilA}
		\tilde{H}_{\mathrm{imp},j} &= U_j (\sigma_0 \otimes \tau_3) + \boldsymbol{J}_j \cdot ( \boldsymbol{\sigma} \otimes \tau_0) ,
		\\
		\label{eq-4by4HamilB}
		\tilde{H}_{\mathrm{elec},\boldsymbol{k}j} &= \sigma_0 \otimes (\xi_{\boldsymbol{k}j} \tau_3 + 
		\Delta_j e^{i \varphi_j \tau_3}\tau_1 ) ,
		\\
		\label{eq-4by4HamilC}
		\tilde{V}_{j} &= v_j (\sigma_0 \otimes \tau_3 ) .
	\end{align}
\end{subequations}
Here, $\xi_{\boldsymbol{k}j}$ is the electronic energy in the superconductor, $\Delta_j$ and $\varphi_j$ are the pairing potential 
and the superconducting phase, respectively, and $v_j$ is the tunnel coupling between the impurity and the superconductor. 
Furthermore, $\sigma_\alpha$ and $\tau_\alpha$ are Pauli matrices $(\alpha \in \{1,2,3\})$ in spin and Nambu space, respectively, 
while $\sigma_0$ and $\tau_0$ are the corresponding unit matrices in these spaces.

To simplify the formalism, it is convenient to transfer the dependence on $\theta_j$ and $\varphi_j$ to the coupling term $V$ 
in Eq.~\eqref{eq-HLR} and work with Hamiltonians describing the subsystems in which the corresponding spin points along its 
quantization axis. Therefore, we introduce the combined unitary transformation $R_j = e^{i \theta_j \sigma_2 / 2} \otimes 
e^{- i \varphi_j \tau_3 / 2}$ in the Hamiltonian defined in Eq.~\eqref{eq-idividualHamilSubsys} to rotate the individual bases 
defined in Eq.~\eqref{eq-basisTot} to the quantization axis in subsystem $j$ along $\boldsymbol{J}_j$ and to remove the phase 
$\varphi_j$. This results in the new bases $\hat{\boldsymbol{d}}_{j} = R_j \tilde{\boldsymbol{d}}_{j}$ and 
$\hat{\boldsymbol{c}}_{\boldsymbol{k}j} = R_j \tilde{\boldsymbol{c}}_{\boldsymbol{k}j}$ and the transformed Hamiltonians
\begin{subequations}
	\begin{align}
	\hat{H}_{\mathrm{imp},j} &= 
	R_j \tilde{H}_{\mathrm{imp},j} R_j^\dag
	= U_j (\sigma_0 \otimes \tau_3) + J_j  ( \sigma_3 \otimes \tau_0) ,
	\\
	\hat{H}_{\mathrm{elec},\boldsymbol{k}j} &= R_j \tilde{H}_{\mathrm{elec},\boldsymbol{k}j}  R_j^\dag
	= \sigma_0 \otimes (\xi_{\boldsymbol{k}j} \tau_3 + \Delta_j \tau_1 ) .
	\end{align}
\end{subequations}

The starting point for the calculation of the electronic transport in the system under study is the calculation of the bare 
Green's function of the impurity coupled to the superconductor in each  subsystem $j$. Following the exact same steps of the 
calculation presented in Ref.~\cite{Villas2020}, we derive the block-diagonal bare matrix Green's function in the new basis 
$\hat{\boldsymbol{d}}_{j}$, i.e.,
\begin{equation}
	\label{eq-gjj}
	\hat g_{jj}(E) =   \begin{pmatrix} \hat g_{jj,\uparrow \uparrow}(E) & 0 \\
	0 & \hat g_{jj,\downarrow \downarrow}(E) \end{pmatrix}   ,
\end{equation}
where the two blocks are given by
\begin{widetext}
	\begin{equation}
	\label{eq-impurity-dressed}
	\hat g_{jj, \sigma \sigma}(E) = \frac{1}{D_{j\sigma}(E)}  \begin{pmatrix}
	E \, \Gamma_{j} + (E+U_j-J_{j\sigma}) \sqrt{\Delta^2_{j} - E^2} & 
	\Gamma_{j} \Delta_{j}   \\
	\Gamma_{j} \Delta_{j}  &
	E \, \Gamma_{j} + (E-U_j-J_{j\sigma}) \sqrt{\Delta^2_{j} - E^2} \end{pmatrix} 
	\end{equation}
\end{widetext}
with the denominator
\begin{multline}
\label{eq-Dup}
D_{j\sigma}(E)  =  2\Gamma_{j} E (E-J_{j\sigma})    \\
+ \left[(E-J_{j\sigma})^2 - U_j^2 -\Gamma^2_{j} \right] \sqrt{\Delta^2_{j} - E^2} .
\end{multline}
Along the derivation, we defined $J_{j\uparrow} = + J_j$, $J_{j\downarrow} = -J_j$ and the tunneling rates $\Gamma_{j} = 
\pi N_{0, j} v^2_{j}$, where $N_{0, j}$ is the normal density of states at the Fermi energy in superconductor $j$.

The current-voltage characteristics of this system will reflect the electronic structure of the magnetic impurities and, in 
particular, the presence of YSR states \cite{Villas2020,Huang2020b}.
From Eqs.~(\ref{eq-gjj}) and (\ref{eq-impurity-dressed}), it follows that the electronic local density of states (LDOS) 
projected onto the impurity site $j$ is given by 
\begin{equation}
\label{eq-LDOS-imp}	
\rho_{{\rm Total},j}(E) = \rho_{j\uparrow}(E) + \rho_{j\downarrow}(E), 
\end{equation}
with
\begin{equation}
\label{eq-LDOS-imp-sigma}
	\rho_{j\sigma}(E) =  \frac{1}{\pi} \mbox{Im} \bigl\{ \hat g_{jj, \sigma \sigma, 11}^\mathrm{a}(E)  \bigr\} ,
\end{equation}
where retarded (r) and advanced (a) Green's functions are defined as $\hat g_{jj, \sigma \sigma, 11}^\mathrm{r,a}(E) = 
\hat g_{jj, \sigma \sigma, 11}(E \pm i \eta_j)$ by introducing the phenomenological Dynes parameter $\eta_j$ which describes 
the inelastic broadening of the electronic states in electrode $j$. The condition for the appearance of superconducting bound 
states is $D_{j\sigma}(E) = 0$. In particular, the spin-induced YSR states appear in the limit $|J_j| \gg \Delta_j$ and they 
are inside the gap when also $\Gamma_j \gg \Delta_j$. In this case, there is a pair of fully spin-polarized YSR bound states 
at energies (measured with respect to the Fermi energy) \cite{Villas2020,Huang2020b}
\begin{equation}
\label{eq-YSR1}
\varepsilon_{j} = \pm \Delta_{j} \frac{J_j^2 - \Gamma^2_{j} - U_j^2}
{\sqrt{ \left[ \Gamma^2_{j} + (J_j-U_j)^2 \right] \left[ \Gamma^2_{j} + (J_j+U_j)^2 \right]}} ,
\end{equation}
which in the electron-hole symmetric case $U_j = 0$ reduces to
\begin{equation}
\label{eq-YSR2}
\varepsilon_{j} = \pm \Delta_{j} \frac{J_j^2 - \Gamma^2_{j}} {J_j^2 + \Gamma^2_{j}} . 
\end{equation}

\subsection{Tunnel coupling between two impurities}

The tunnel coupling $V$ in Eq.~\eqref{eq-HLR} between the two subsystems with the global quantization axis defined by 
Eq.~\eqref{eq-basis} reads
\begin{align}
	V &= \frac{1}{2}  \tilde{\boldsymbol{d}}_\mathrm{t}^\dag \tilde{V}_\mathrm{tS} \tilde{\boldsymbol{d}}_\mathrm{S}^{\phantom \dag} 
	+ \frac{1}{2}  \tilde{\boldsymbol{d}}_\mathrm{S}^{ \dag}  \tilde{V}_\mathrm{St} \tilde{\boldsymbol{d}}_\mathrm{t}^{\phantom \dag} ,
\end{align}
with $\tilde{V}_\mathrm{St} = v (\sigma_0 \otimes \tau_3) = \tilde{V}_\mathrm{tS}^\dag$ and the tunnel coupling $v$ between the 
two impurities \cite{Villas2020}. Introducing the aforementioned basis rotation $R_j$ in subsystem $j$ results in
\begin{align}
	V &= \frac{1}{2} \hat{\boldsymbol{d}}_\mathrm{t}^\dag \hat{V}_\mathrm{tS} \hat{\boldsymbol{d}}_\mathrm{S} 
		+ \frac{1}{2} \hat{\boldsymbol{d}}_\mathrm{S}^\dag \hat{V}_\mathrm{St}  \hat{\boldsymbol{d}}_\mathrm{t}^{\phantom \dag}
\end{align}
where
\begin{subequations}
	\label{eq-rotatedHopping}
	\begin{align}
	\hat{V}_\mathrm{tS} &= 
	R_\mathrm{t}
	\tilde{V}_\mathrm{tS} 
	R_\mathrm{S}^\dag
	=
	v (e^{-i \theta \sigma_2 / 2} \otimes \tau_3 e^{- i \varphi_0 \tau_3 / 2} ) ,
	\\
	\hat{V}_\mathrm{St}  &=
	R_\mathrm{S} 
	\tilde{V}_\mathrm{St} 
	R_\mathrm{t}^\dag 
	=
	v (e^{i \theta \sigma_2 / 2} \otimes \tau_3  e^{ i \varphi_0 \tau_3 / 2} ) ,
	\end{align}
\end{subequations}
$\theta = \theta_\mathrm{S} - \theta_\mathrm{t}$ is the relative angle, and $\varphi_0 = \varphi_\mathrm{t} - \varphi_\mathrm{S}$ 
the superconducting phase difference between the two impurities. In that sense, the coupling between the two subsystems is 
effectively represented by a spin-active interface in which there are spin-flip processes whose probabilities depend on the 
relative orientation of the impurity spins described by $\theta$.

\subsection{Calculation of the current-voltage characteristics}

To compute the electronic transport properties in our model system, we shall assume that the voltage drops at the interface 
between the two impurities, which is justified by the fact that usually the impurity-impurity coupling $v$ is much weaker than 
the impurity-electrode couplings $v_j$. Under this assumption, our system effectively reduces to a superconducting quantum point 
contact and we can compute its transport properties with a generalization of the MAR theory of Ref.~\cite{Cuevas1996} to account 
for the spin-dependent transport. This generalization was in fact developed in our previous work of Ref.~\cite{Villas2020} and we 
simply reproduce the formalism here to make this manuscript more self-contained and to emphasize the peculiarities introduced by 
the spin-flip processes between the two impurities.

Our goal is to compute the current in our two-impurity system under an external bias voltage $V$. As in any superconducting contact, 
the bias voltage induces a time-dependent superconducting phase difference $\varphi(t) = \varphi_0 + 2eVt/\hbar$ that varies linearly 
in time with the bias. This can be simply included in the formalism by replacing $\varphi_0$ with $\varphi(t)$ in 
Eq.~\eqref{eq-rotatedHopping} such that $\hat{V}_{jk}$ acquires a time dependence $\hat{V}_{jk}(t)$. The theory of Ref.~\cite{Cuevas1996} 
is based on nonequilibrium Green's function techniques (or Keldysh formalism) and a central role is played by the lesser $4 \times 4$ 
matrix Green's functions 
\begin{equation}
\hat G^{+-}_{jk}(t,t^{\prime}) = -i \langle T_\mathrm{C} \bigl\{ 
\hat{\boldsymbol{d}}_j(t_{+}) \otimes \hat{\boldsymbol{d}}^{\dagger}_k(t^{\prime}_{-}) \bigr\} \rangle ,
\end{equation}
for $j,k \in \{ \mathrm{t, S} \}$ and where $\hat{\boldsymbol{d}}_j$ and $\hat{\boldsymbol{d}}^{\dagger}_k$  are the rotated 
four-component spinors defined above. In addition, $T_\mathrm{C}$ is the time-ordering operator on the Keldysh contour such that 
any time in the lower branch ($t^{\prime}_{-}$) is larger than any time in the upper one ($t_{+}$). The electrical current in our 
system is defined as $I(t) = - e \langle\mathrm{d} N_\mathrm{S}(t) / \mathrm{d} t \rangle$, where $N_\mathrm{S} = \sum_{\sigma} 
d_{\mathrm{S}\sigma}^\dag d_{\mathrm{S}\sigma}^{\phantom \dag}$ is the number operator in subsystem S, and it can be expressed in 
terms of $\hat G^{+-}_{jk}$ as \cite{Villas2020} 
\begin{multline}
	\label{eq-current1}
	I(t)  =  \frac{e}{2\hbar} \mbox{Tr} \Bigl\{ (\sigma_0 \otimes \tau_3) \Bigl[ 
	\hat{V}_\mathrm{St}(t) \hat G^{+-}_\mathrm{tS}(t,t)    \\ 
	 -  \hat{V}_\mathrm{tS}(t) 
	\hat G^{+-}_\mathrm{St}(t,t) \Bigr] \Bigr\}, 
\end{multline}
where $\mbox{Tr}$ is the trace taken over spin and Nambu degrees of freedom.

The task is now to compute the dressed Green's functions $G^{+-}_{jk}$ appearing in the current formula. For this purpose, we follow 
a perturbative scheme and treat the coupling term in the Hamiltonian of Eq.~(\ref{eq-HLR}) as a perturbation. The unperturbed Green's 
functions $\hat g_{jj}$ correspond to the uncoupled impurity-electrode subsystems $j$ in equilibrium and are given by Eq.~\eqref{eq-gjj}. 
On the other hand, to solve the problem it is convenient to express the current in terms of the so-called $T$-matrix. The $T$-matrix 
associated with the time-dependent perturbation is defined as
\begin{equation}
	\hat{T}^\mathrm{r,a} = \hat{V} + \hat{V} \circ \hat{g}^\mathrm{r,a} \circ \hat{T}^\mathrm{r,a} ,
\end{equation}
where the $\circ$ product is a shorthand for convolution, i.e., for integration over intermediate time arguments. As shown in 
Ref.~\cite{Cuevas1996}, the exact current to all orders in the tunneling rate can be written in terms of the $T$-matrix components as
\begin{widetext}
\begin{align}
I(t)  =  \frac{e}{2\hbar} 	\mbox{Tr} \Bigl\{ (\sigma_0 \otimes \tau_3) & \Bigl[
\hat T^\mathrm{r}_\mathrm{St} \circ \hat g^{+-}_\mathrm{tt} \circ 
\hat T^\mathrm{a}_\mathrm{tS} \circ \hat g^\mathrm{a}_\mathrm{SS}
- \hat g^\mathrm{r}_\mathrm{SS} \circ \hat T^\mathrm{r}_\mathrm{St} \circ 
\hat g^{+-}_\mathrm{tt} \circ \hat T^\mathrm{a}_\mathrm{tS}   
\nonumber \\ 
 & +
\hat g^\mathrm{r}_\mathrm{tt} \circ \hat T^\mathrm{r}_\mathrm{tS} \circ 
\hat g^{+-}_\mathrm{SS} \circ \hat T^\mathrm{a}_\mathrm{St} - 
\hat T^\mathrm{r}_\mathrm{tS} \circ \hat g^{+-}_\mathrm{SS} \circ 
\hat T^\mathrm{a}_\mathrm{St} \circ \hat g^\mathrm{a}_\mathrm{tt}
\Bigr] \Bigr\} .
\end{align}
\end{widetext}
It is convenient to Fourier transform with respect to the temporal arguments to solve the $T$-matrix integral equations: 
\begin{equation}
\hat T(t,t^{\prime}) = \frac{1}{2\pi} \int^{\infty}_{-\infty} dE \int^{\infty}_{-\infty} dE^{\prime} 
e^{-iEt} e ^{iE^{\prime} t^{\prime}} \hat T(E,E^{\prime}) .	
\end{equation}
Because of the time dependence of the coupling matrices, one can show that $\hat T(E,E^{\prime})$ admits the following 
general solution
\begin{equation}
\hat T(E,E^{\prime}) = \sum_n \hat T(E, E +neV) \delta(E-E^{\prime} + neV) .
\end{equation}
Thus, it follows that the current exhibits a time dependence in the form of the Fourier series 
\begin{equation}
\label{eq-It}
I(t) = \sum_n I_n e^{i n \varphi(t)} ,	
\end{equation}
where the current amplitudes $I_n$ can be expressed in terms of the components $\hat T_{nm}(E) = \hat T(E+neV,E+meV)$ and 
$\hat g_{jj,n}(E) = \hat g_{jj}(E+neV)$ as
\begin{widetext}
\begin{align}
\label{I_full}
I_n =  \frac{e}{2h} \int^{\infty}_{-\infty} dE \sum_m 
\mbox{Tr} \Bigl\{ (\sigma_0 \otimes \tau_3) & \Bigl[	
\hat T^\mathrm{r}_{\mathrm{St},0m} \hat g^{+-}_{\mathrm{tt},m} 
\hat T^\mathrm{a}_{\mathrm{tS},mn} \hat g^\mathrm{a}_{\mathrm{SS},n}
- \hat g^\mathrm{r}_{\mathrm{SS},0} \hat T^\mathrm{r}_{\mathrm{St},0m} 
\hat g^{+-}_{\mathrm{tt},m} \hat T^\mathrm{a}_{\mathrm{tS},mn} 
 \nonumber \\ 
&  +
\hat g^\mathrm{r}_{\mathrm{tt},0} \hat T^\mathrm{r}_{\mathrm{tS},0m} 
\hat g^{+-}_{\mathrm{SS},m} \hat T^\mathrm{a}_{\mathrm{St},mn} - 
\hat T^\mathrm{r}_{\mathrm{tS},0m} \hat g^{+-}_{\mathrm{SS},m} 
\hat T^\mathrm{a}_{\mathrm{St},mn} \hat g^\mathrm{a}_{\mathrm{tt},n}
\Bigr] \Bigr\} .
\end{align}
\end{widetext}
Notice that the bare Green's functions are diagonal in energy space and the bare lesser Green's functions are given by 
$\hat g^{+-}_{jj}(E) = \left[ \hat g^\mathrm{a}_{jj}(E) - \hat g^\mathrm{r}_{jj}(E) \right] f(E)$, where $f(E) = \left[ 1 + 
\exp(E/k_\mathrm{B}T) \right]^{-1}$ is the Fermi function with temperature $T$ and the Boltzmann constant $k_\mathrm{B}$. 
The previous formula can be further simplified by using the general relation $\hat T^\mathrm{r,a}_{\mathrm{tS},nm}(E) = 
(\hat T^\mathrm{a,r}_{\mathrm{St},mn})^{\dagger}(E)$, which reduces the calculation of the current to the determination of the 
Fourier components $\hat T^\mathrm{r,a}_{\mathrm{St},nm}$ fulfilling the set of linear algebraic equations
\begin{multline}
\label{eq-T}
\hat T^\mathrm{r,a}_{\mathrm{St},nm} = \hat{V}_{\mathrm{St},nm} + 
\hat {\cal E}^\mathrm{r,a}_n \hat T^\mathrm{r,a}_{\mathrm{St},nm} \\ 
+ \hat {\cal W}^\mathrm{r,a}_{n,n-2} \hat T^\mathrm{r,a}_{\mathrm{St},n-2,m} + 
\hat {\cal W}^\mathrm{r,a}_{n,n+2} \hat T^\mathrm{r,a}_{\mathrm{St},n+2,m} ,
\end{multline}
where the different matrix coefficients are given in terms of the unperturbed Green's functions as
\begin{subequations}
	\begin{align} 
	\label{eq-coeff}
		\hat V_{\mathrm{St},nm} &=  \frac{v}{2} e^{i \theta \sigma_2 / 2} 
		\otimes  \bigl[ (\tau_3 + \tau_0) \delta_{n+1,m} 
		\nonumber \\ 
		&\qquad \qquad \qquad \qquad + (\tau_3 - \tau_0) \delta_{n-1,m} \bigr] , 
		 \\
		 \hat V_{\mathrm{tS},nm} &=  \frac{v}{2} e^{-i \theta \sigma_2 / 2} 
		 \otimes  \bigl[ (\tau_3 + \tau_0) \delta_{n-1,m} 
		 \nonumber \\ 
		 &\qquad \qquad \qquad \qquad + (\tau_3 - \tau_0) \delta_{n+1,m} \bigr] , 
		 \\
		 \hat{\mathcal{E}}_{n}^{\mathrm{r,a}} &= 
		 \bigl[ \hat{V}_{\mathrm{St},n,n+1} \,
		 \hat{g}_{\mathrm{tt},n+1}^{\mathrm{r,a}} \,
		 \hat{V}_{\mathrm{tS},n+1,n} 
		 \nonumber \\
		 & \qquad + \hat{V}_{\mathrm{St},n,n-1} \,
		 \hat{g}_{\mathrm{tt},n-1}^{\mathrm{r,a}} \,
		 \hat{V}_{\mathrm{tS},n-1,n} 
		 \bigr]
		 \hat{g}_{\mathrm{SS},n}^{\mathrm{r,a}} ,
		 \\
		 \hat{\mathcal{W}}_{n,n-2}^\mathrm{r,a} &=
		 \hat{V}_{\mathrm{St},n,n-1} \,
		 \hat{g}_{\mathrm{tt},n-1}^{\mathrm{r,a}} \,
		 \hat{V}_{\mathrm{tS},n-1,n-2} \,
		 \hat{g}_{\mathrm{SS},n-2}^{\mathrm{r,a}} ,
		 \\
		 \hat{\mathcal{W}}_{n,n+2}^\mathrm{r,a} &=
		 \hat{V}_{\mathrm{St},n,n+1} \,
		 \hat{g}_{\mathrm{tt},n+1}^{\mathrm{r,a}} \,
		 \hat{V}_{\mathrm{tS},n+1,n+2} \,
		 \hat{g}_{\mathrm{SS},n+2}^{\mathrm{r,a}} .
	 \end{align}
\end{subequations}
In general, these block-tridiagonal systems have to be solved numerically and the current can only be expressed in an analytical form 
in the tunnel regime, as we discuss in Sec.~\ref{sec-tunnel-regime}. On the other hand, let us stress that we shall focus here exclusively 
on the discussion of the dc current, i.e., $I_0$ in Eq.~(\ref{eq-It}), and we shall not analyze the (zero-bias) dc Josephson current (or 
supercurrent).

\subsection{Normal state conductance} \label{sec-GN}

To get insight into the current in our system, it is didactic to consider the case in which the electrodes are in the normal state. 
Moreover, the analysis of this case gives us the chance to introduce the normal state conductance, $G_{\rm N}$, which is the physical 
parameter that allows to make contact with the experiment. In the case in which neither the tip nor the substrate are superconducting, 
the current formula within our model can be worked out analytically and it is given by the following Landauer-type of expression
\begin{multline}
	I_{\rm normal}(V, \theta) = \frac{e}{h} \sum_{\sigma, \sigma^{\prime}} \int^{\infty}_{-\infty} dE \, 
		\tau_{\sigma, \sigma^{\prime}}(E,V,\theta) \\ 
		\times \left[ f(E-eV) - f(E) \right] , 
\end{multline}
where $\tau_{\sigma, \sigma^{\prime}}(E,V,\theta)$ are the transmission coefficients for electron tunneling processes connecting
spins $\sigma$ and $\sigma^{\prime}$. In general, the expressions of these coefficients in terms of the different parameters of
the model are extremely cumbersome and in what follows, we only provide such expressions in certain limiting cases. First of all, in
the tunnel regime, where $v \ll \Gamma_\mathrm{t,S}$, we find
\begin{widetext}
\begin{subequations}
	\begin{align} 
	   \tau_{\sigma, \sigma}(E,V,\theta) & \approx \frac{4v^2 \Gamma_\mathrm{S} \Gamma_\mathrm{t} \cos^2 (\theta/2)}
	   {[(E-eV - U_\mathrm{S} - J_{\mathrm{S}\sigma})^2 + \Gamma^2_\mathrm{S}] 
	    [(E - U_\mathrm{t} - J_{\mathrm{t}\sigma})^2 + \Gamma^2_\mathrm{t}]} , \\
	    \tau_{\sigma, \bar \sigma}(E,V,\theta) & \approx \frac{4v^2 \Gamma_\mathrm{S} \Gamma_\mathrm{t} \sin^2 (\theta/2)}
	   {[(E-eV - U_\mathrm{S} - J_{\mathrm{S}\sigma})^2 + \Gamma^2_\mathrm{S}] 
	    [(E - U_\mathrm{t} + J_{\mathrm{t}\sigma})^2 + \Gamma^2_\mathrm{t}]} ,	
	\end{align}
\end{subequations}
where $\bar \sigma = - \sigma$. Notice that, as expected, the coefficient for antiparallel spins vanishes when $\theta=0$.
Moreover, in the limit in which we are interested, namely the limit when YSR states appear, one can safely ignore the energy
and bias dependence of these transmission coefficients. On the other hand, and to give an idea about these coefficients beyond 
the tunnel regime, we consider the case of parallel spin ($\theta = 0$). In this case (ignoring the bias dependence),
\begin{subequations}
	\begin{align} 
	\label{eq-taus-tunnel}
	   \tau_{\sigma, \sigma}(E,0,0) & = \frac{4v^2 \Gamma_\mathrm{S} \Gamma_\mathrm{t}}
	   {[(E - U_\mathrm{S} - J_{\mathrm{S}\sigma})^2 + \Gamma^2_\mathrm{S}] 
	    [(E - U_\mathrm{t} - J_{\mathrm{t}\sigma})^2 + \Gamma^2_\mathrm{t}] -
	    2 v^2[(E - U_\mathrm{S} - J_{\mathrm{S} \sigma}) ( E - U_\mathrm{t} - J_{\mathrm{t} \sigma})- 
	    \Gamma_\mathrm{S} \Gamma_\mathrm{t} ] + v^4} , \\
	    \tau_{\sigma, \bar \sigma}(E,0,0) & = 0  .	
	\end{align}
\end{subequations}
\end{widetext}
	
In general, the zero-temperature normal state linear conductance in our system is given by
\begin{equation}
\label{eq-GN}
	\frac{G_{\rm N}}{G_0} =  \frac{1}{2} \sum_{\sigma, \sigma^{\prime}} 
	\tau_{\sigma, \sigma^{\prime}}(E=0,V=0,\theta)  ,
\end{equation}
where $G_0 = 2e^2/h$ is the quantum of conductance. Moreover, in this work, $|eV|$ will always be much smaller than
$\Gamma_{\rm t} + \Gamma_{\rm S}$ such that the differential conductance in the normal state will be independent of the bias.

\section{Tunnel regime} \label{sec-tunnel-regime}

So far, the experiments on the tunneling between YSR states have been performed in the so-called tunnel regime, in which the 
coupling between the impurities is relatively weak and the only transport process that takes place is single-quasiparticle 
tunneling (eventually involving the YSR states) \cite{Huang2020a}. This regime has already been addressed in 
Refs.~\cite{Huang2020a,Huang2020c} and we want to expand that discussion in this section in the light of the model described 
in the previous section.

Let us recall that the main experimental observation reported in Ref.~\cite{Huang2020a} is the appearance of current peaks inside 
the gap region that can be associated with the quasiparticle tunneling between the YSR states of the two impurities. Let us now show 
how this observation can be naturally explained within our model. In our case, the tunnel regime can be roughly defined as the limit 
in which the tunnel coupling is sufficiently weak such that $v^2 \ll \Gamma_{\rm S} \Gamma_{\rm t}$ and the only relevant tunneling 
process is the single-quasiparticle tunneling. In this limit, we can use the approximation $\hat T^\mathrm{r,a}_{\mathrm{St},nm} 
\approx \hat V_{\mathrm{St},nm}$ in Eq.~\eqref{eq-T} and after some straightforward algebra we arrive at the following expression 
for the tunneling current at the lowest order in the tunnel coupling between the impurities
\begin{eqnarray}
\label{eq-I_tunnel}
{I}(V,\theta) & = & \frac{4 \pi^2 e v^2}{h}  \sum_{\sigma} \int^{\infty}_{-\infty} dE \, 
\left[ f(E-eV) - f(E) \right] \nonumber \\ & & \times \left\{
\cos^2 (\theta/2) \rho_{\rm S\sigma}(E-eV) \rho_{\rm t\sigma}(E) \right. \nonumber \\
& & \hspace{2mm} + \left. \sin^2 (\theta/2) \rho_{\rm S\sigma}(E-eV) \rho_{\rm t\bar{\sigma}}(E) \right\} .
\end{eqnarray}
Let us recall that in this expression $v$ is the hopping element that describes the coupling between the impurities, $f(E)$ is the 
Fermi function, $\theta$ is the angle defining the relative orientation of the impurity spins, and $\rho_{j \sigma}$ is the LDOS 
on the impurity site $j=\mathrm{t,S}$ for spin $\sigma$ ($\bar \sigma$ stands for the spin antiparallel to $\sigma$), which is 
given by Eq.~(\ref{eq-LDOS-imp-sigma}). The current formula of Eq.~(\ref{eq-I_tunnel}) has the expected structure for a tunnel 
junction with a spin-active interface. As usual in those junctions, we have two types of processes: (i) tunnel events involving parallel 
spins (terms weighted by $\cos^2(\theta/2)$) and (ii) tunnel events involving antiparallel spins (terms weighted by $\sin^2(\theta/2)$).
When both electrodes are in the normal states, this result reduces to that described in Sec.~\ref{sec-GN} for the tunnel regime.

\begin{figure*}[t]
\includegraphics[width=0.9\textwidth,clip]{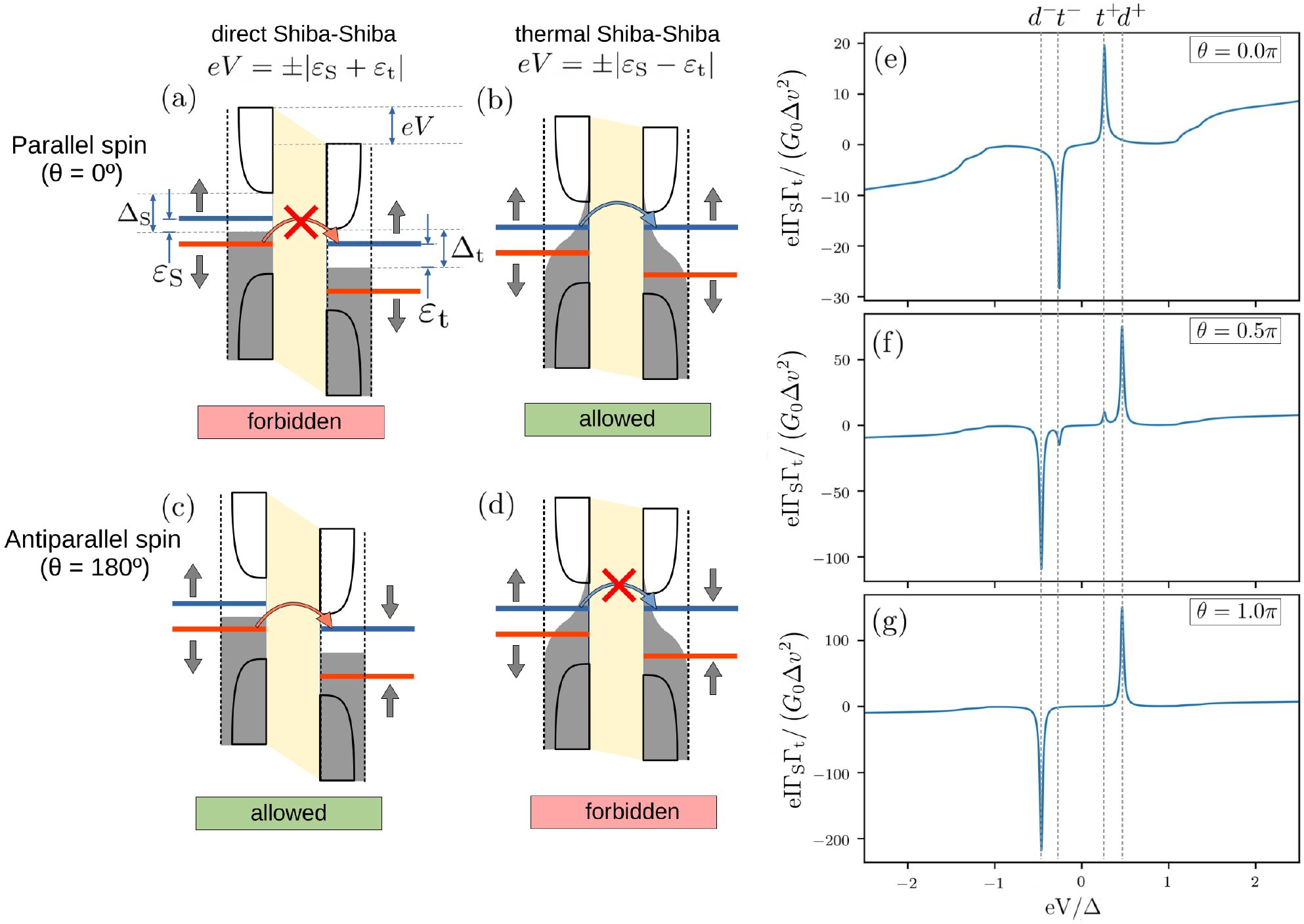}
\caption{(a) Tunnel process corresponding to direct Shiba-Shiba tunneling that is forbidden in the case of parallel spins.
(b) Tunnel process corresponding to the thermally activated Shiba-Shiba tunneling that is allowed in the case of parallel spins.
(c) Direct Shiba-Shiba tunneling that is allowed in the case of antiparallel spins. 
(d) Thermal Shiba-Shiba tunneling that is forbidden in the case of antiparellel spins. 
(e) Current-voltage characteristics in the tunnel regime for collinear spins ($\theta = 0$), 
as computed from Eq.~(\ref{eq-I_tunnel}), and for $\Delta_{\rm S} = \Delta_{\rm t} = \Delta$, $\Gamma_{\rm S} = \Gamma_{\rm t} = 
100 \Delta$, $U_{\rm S} = 0$, $J_{\rm S} = 90 \Delta$, $U_{\rm t} = 20 \Delta$, $J_{\rm t} = 70 \Delta$, $\eta_\mathrm{S} = 
\eta_\mathrm{t} = 0.01\Delta$, $v = \Delta$, and $k_\mathrm{B}T = 0.05\Delta$. (f) The same as in panel (e), but for 
$\theta = 0.5 \pi$. (g) The same as in panel (e), but for antiparallel spins ($\theta = \pi$). The vertical dotted lines in panels 
(e-g) indicate the expected energies of the current peaks originating from the direct Shiba-Shiba tunneling ($d^+$ and $d^-$), 
$\pm |\varepsilon_\mathrm{S} + \varepsilon_{t}| = \pm 0.47\Delta$, and from the thermal Shiba-Shiba tunneling ($t^+$ and $t^-$), 
$\pm |\varepsilon_\mathrm{S} - \varepsilon_{t}| = \pm 0.26\Delta$. Notice that for $\theta = 0$ only the thermal Shiba-Shiba peaks 
are observed, for $\theta = \pi$ only the direct Shiba-Shiba peaks show up, and for $\theta = \pi/2$ both types of current peaks 
are visible in the subgap region.} 
\label{fig-tunnel1}
\end{figure*}

In Fig.~\ref{fig-tunnel1} we illustrate the results obtained with the tunnel formula above for three different values of the angle 
$\theta$ together with a schematic description of the processes. In this example, as in all cases discussed in this manuscript, we assume 
equal superconducting gaps for the tip and the substrate $\Delta_\mathrm{S} = \Delta_\mathrm{t} = \Delta$ and set $\Gamma_\mathrm{S} =
\Gamma_\mathrm{t} = 100\Delta$ (to be in the strong coupling regime realized in STM experiments in which YSR states appear). Additionally, 
we have $U_\mathrm{S} = 0$ and $J_\mathrm{S} = 90\Delta$ for the impurity coupled to the substrate, and $U_\mathrm{t} = 20\Delta$ and 
$J_\mathrm{t} = 70\Delta$ for the impurity coupled to the tip (the large values of $J$, comparable to $\Gamma_\mathrm{S,t}$, 
are necessary for the YSR states to be well inside the gap). With these parameter values, the YSR states in both impurities appear 
at energies $\pm \varepsilon_\mathrm{S} = \pm 0.105\Delta$ and $\pm \varepsilon_\mathrm{t} = \pm 0.365\Delta$ (we assume that 
$\varepsilon_\mathrm{t,S} > 0$). Finally, we have assumed a finite temperature of $k_\mathrm{B}T = 0.05\Delta$. 

The result for parallel spins ($\theta = 0$) is shown in panel Fig.~\ref{fig-tunnel1}(e). In this case, the most salient feature is 
the appearance of two current peaks inside the gap region at a bias $eV = \pm |\varepsilon_\mathrm{S} - \varepsilon_\mathrm{t}| = 
\pm  0.26\Delta$. Since in this case the impurity spins are parallel, the tunneling between the lower YSR state in one impurity and 
the upper state in the other impurity is forbidden, as we illustrate in Fig.~\ref{fig-tunnel1}(a). Notice that in this example both 
impurities have the same type of ground state, i.e., they are on the same side of the quantum critical point (the point in parameter
space for which the YSR states appear at zero energy and the spin of the ground state changes). Thus, a subgap current peak in the 
tunnel regime for $\theta = 0$ can only be due to the tunneling between the two upper (or two lower) states, which is possible due 
to the finite temperature and the corresponding partial occupation of the different states, see Fig.~\ref{fig-tunnel1}(b). For this 
reason, we refer to these peaks as \emph{thermal Shiba-Shiba peaks} and denote their height as $t^+$ and $t^{-}$ for positive ($+$) 
and negative ($-$) bias. Notice that in this case $t^+ \neq t^-$ because of the lack of electron-hole symmetry in the tip impurity 
($U_\mathrm{t} \neq 0$).

Let us now discuss the case of antiparallel spins ($\theta = \pi$) shown in Fig.~\ref{fig-tunnel1}(g). In this case, the
tunneling between the lower and upper YSR states is allowed, see Fig.~\ref{fig-tunnel1}(c), and this process gives rise
to current bias at $eV = \pm |\varepsilon_\mathrm{S} + \varepsilon_\mathrm{t}| = \pm 0.47\Delta$, which explains the subgap 
structure shown in Fig.~\ref{fig-tunnel1}(g). We refer to the peaks originating from this tunneling process as \emph{direct 
Shiba-Shiba peaks} and we denote their height as $d^+$ and $d^{-}$ for positive ($+$) and negative ($-$) bias. Again, the
fact that $d^+ \neq d^-$ in this example is due to the electron-hole asymmetry in the tip impurity. In the case of
antiparallel spins ($\theta = \pi$), the thermally activated processes described in the previous paragraph are forbidden,
see Fig.~\ref{fig-tunnel1}(d), which explains the absence of the corresponding peaks at $eV = \pm |\varepsilon_\mathrm{S} - 
\varepsilon_\mathrm{t}| = \pm 0.26\Delta$, see Fig.~\ref{fig-tunnel1}(g). 

For an intermediate situation, when the impurity spins are neither parallel nor antiparallel, both types of processes, direct
and thermal Shiba-Shiba tunneling, are possible and both types of current peaks appear simultaneously at a finite temperature.
This is illustrated in Fig.~\ref{fig-tunnel1}(f) where we show the result for $\theta = \pi/2$. 

Let us recall that in the experiments of Refs.~\cite{Huang2020a,Huang2020c}, both types of peaks were observed at sufficiently
high temperatures, which was interpreted as a sign that the spins were neither parallel nor antiparallel. Actually, the 
detailed analysis presented in Ref.~\cite{Huang2020c} suggested that there was no magnetic anisotropy fixing the relative spin
orientation and that the spins in that experiment were freely rotating. In that case, the current measured in practice is an average 
over all possible values of the angle $\theta$, which can be trivially done from Eq.~(\ref{eq-I_tunnel}) using $\langle \cos^2(\theta/2) 
\rangle = \langle \sin^2(\theta/2) \rangle = 1/2$, where $\langle \ \cdot \ \rangle$ denotes the angular average. The averaged current 
turns out to be equal to the current given by Eq.~(\ref{eq-I_tunnel}) for $\theta = \pi/2$. Thus, the example of 
Fig.~\ref{fig-tunnel1}(f) describes precisely this averaged current in a situation where $\theta$ varies rapidly in time. 

An important finding of Ref.~\cite{Huang2020c} was that the relative orientation between the impurity spins, i.e., the angle $\theta$, 
can be extracted from the ratio between the thermal and the direct Shiba-Shiba peak. This conclusion was drawn with the help of
the classical Shiba model \cite{Shiba1968} and our goal now is to show that it can also be derived from the Anderson model
used in this work. To obtain the height of the different current peaks we first need analytical expressions for the LDOS
describing the YSR states. From Eq.~(\ref{eq-LDOS-imp-sigma}), it is easy to show that the spin-dependent impurity LDOS
for energies close to the bound states adopt a Lorentzian-like form given by
\begin{equation}
\label{eq-Lorentzian}
\rho_{j \sigma}(E) = \frac{1}{\pi} \frac{A_{j \sigma}}{(E-\varepsilon_j)^2 + \eta^2_j} ,
\end{equation}
where $A_{j \sigma}$ is a positive constant and $\eta_j$ describes the broadening (or inverse lifetime) of the corresponding 
bound state in impurity $j=\mathrm{t,S}$. The constants $A_{j \sigma}$ depend on the different parameters of the model, but the 
corresponding expressions are not important for our discussion here. Substituting Eq.~\eqref{eq-Lorentzian} into the current formula of 
Eq.~(\ref{eq-I_tunnel}), we can compute the height of the different peaks. Of importance here is the ratio $r = \sqrt{t^+ t^- /(d^+ d^-)}$ 
involving the height of the four different peaks, thermal and direct for positive and negative bias. It is straightforward to show that
in the limit in which $k_\mathrm{B} T \gg \eta_j$, which is almost always the case even for very low temperatures, this ratio 
is given by
\begin{equation}
r = \sqrt{\frac{t^+ t^-}{d^+ d^-}} = \cot^2\biggl( \frac{\theta}{2} \biggr) \left \vert \frac{f(\varepsilon_{\rm S}) - f(\varepsilon_{\rm t})}
{f(\varepsilon_{\rm S}) - f(-\varepsilon_{\rm t})} \right \vert ,
\end{equation}
which is the result derived in Ref.~\cite{Huang2020c}. Moreover, if $k_\mathrm{B} T \ll \varepsilon_j$, which is often
the case, the previous formula reduces to
\begin{equation}
r = \cot^2\biggl( \frac{\theta}{2} \biggr) \, | e^{-\varepsilon_{\rm S}/k_\mathrm{B}T} - e^{-\varepsilon_{\rm t}/k_\mathrm{B}T} | .
\end{equation}

As explained in Ref.~\cite{Huang2020c}, the importance of this result is that the relative orientation between the impurity 
spins can be obtained from quantities (the current peak heights and the temperature) that can be directly measured. Here, we show 
that this result is quite universal and it does not depend on the details of the impurity model, as long as electron correlations 
can be ignored. 

\begin{figure}[t]
\includegraphics[width=0.9\columnwidth,clip]{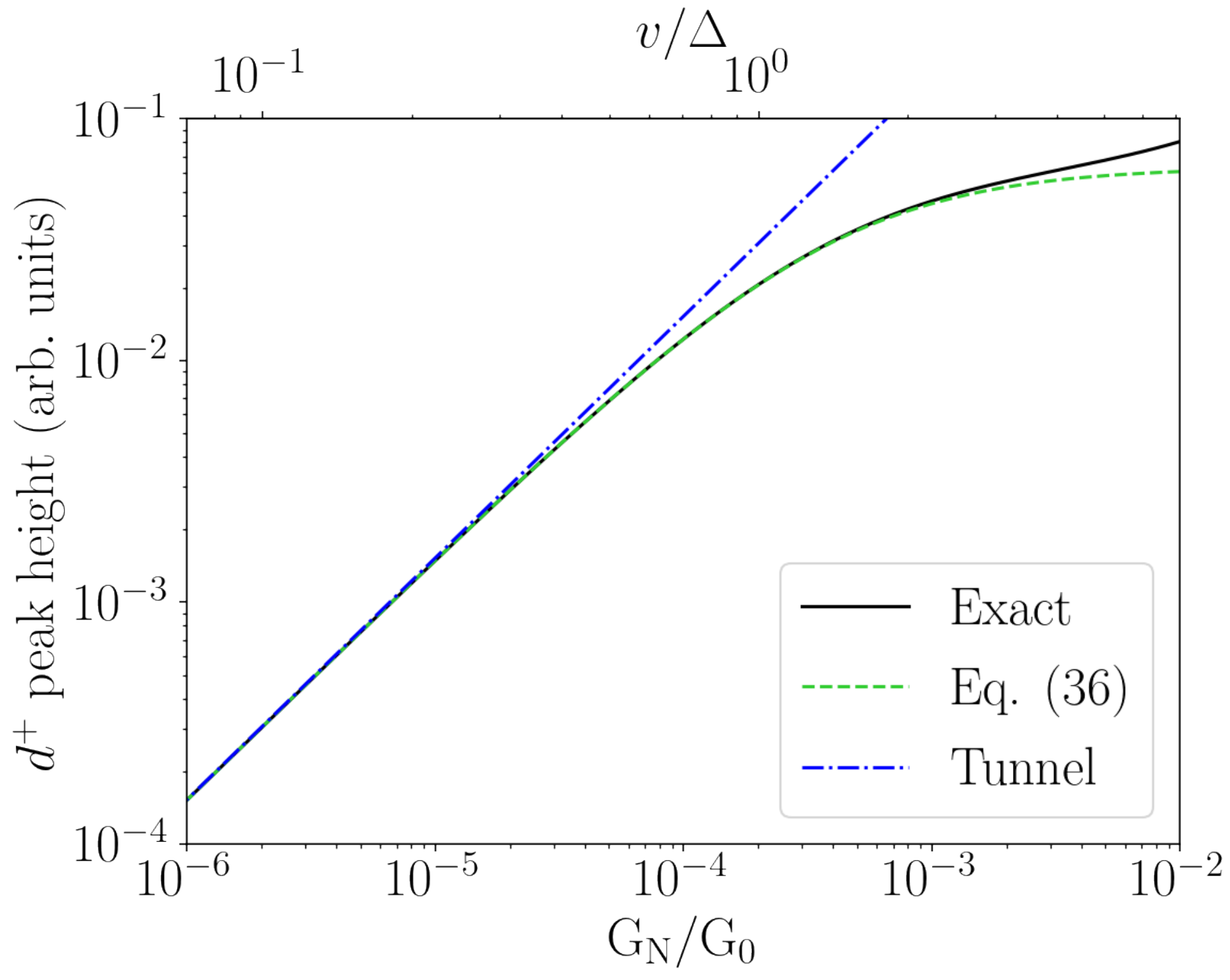}
\caption{Height of the direct Shiba-Shiba peak for positive bias, $d^+$, as a function of the normal state conductance 
$G_\mathrm{N}$, normalized by the quantum of conductance $G_0 = 2e^2/h$. The corresponding values of the tunnel coupling $v$ are 
also shown in the upper horizontal axis. The values of the different model parameters are: $\Delta_{\rm S} = \Delta_{\rm t} = \Delta$, 
$\Gamma_{\rm S} = \Gamma_{\rm t} = 100 \Delta$, $U_{\rm S} = 60 \Delta$, $U_{\rm t} = 0$, $J_{\rm S} = J_{\rm t} = 60 \Delta$, 
$k_\mathrm{B}T = 0$, $\eta_\mathrm{t} = \eta_\mathrm{S} = 0.01\Delta$, and $\theta = \pi$. The dotted-dashed line corresponds to 
the tunnel approximation of Eq.~(\ref{eq-I_tunnel}), the dashed line to the approximation of Eq.~(\ref{eq-I_tunnel_den}), and the 
solid line to the exact result.}
\label{fig-tunnel2}
\end{figure}

Another interesting observation reported in Ref.~\cite{Huang2020a} is the fact that the height of the peaks (and their area) undergoes
a crossover between a linear regime at very low transmission (or normal state conductance) and a sublinear regime at higher 
transmission when the STM tip with its impurity was brought closer to the impurity on the substrate. Obviously, the tunnel
approximation of Eq.~(\ref{eq-I_tunnel}) can only explain the linear regime in which the current, including the current peak
heights, is proportional to $v^2$ and, in turn, to the normal state conductance. This perturbative result must fail at some point 
upon increasing the tunnel coupling, or reducing the bound state broadening, because $v^2$ times the product of density of states
is no longer a small parameter. This has nothing to do with the occurrence of MARs, which were negligible in the experiments of
Ref.~\cite{Huang2020a}. Thus, in order to describe the crossover to a sublinear regime, we must take into account the multiple normal 
reflections that may take place in the resonant electron tunneling between two sharp bound states (as in any resonant tunneling
situation). In our case, this can be achieved by neglecting the anomalous Green's function in the $T$-matrix equations, which amounts
to ignore the Andreev reflections, and solving them to infinite order in the tunnel coupling. Technically speaking, this is
done by approximating Eq.~(\ref{eq-T}) by 
\begin{eqnarray}
\label{eq-T-approx}
\hat T^\mathrm{r,a}_{\mathrm{St},nm} & = & 
\left[\hat 1 - \hat {\cal E}^\mathrm{r,a}_n\right]^{-1} \hat V_{\mathrm{St},nm} ,
\end{eqnarray}
where, in addition, the anomalous Green's functions (off-diagonal components in Nambu space) are set to zero in the expression of 
$\hat {\cal E}^\mathrm{r,a}_n$. Then, the solution of this equation can be introduced in the current formula of Eq.~(\ref{I_full}).
Finally, after some algebra and retaining only the lowest order terms in $v$ in the numerator, we arrive at the following 
improved formula for the tunneling current
\begin{eqnarray}
\label{eq-I_tunnel_den}
{I}(V,\theta) & = & \frac{4 \pi^2 e v^2}{h}  \sum_{\sigma} \int^{\infty}_{-\infty} dE \, 
\frac{\left[ f(E-eV) - f(E) \right]}{ \vert \tilde D(E) \vert^2} 
\nonumber \\ & & \times \left\{
\cos^2 (\theta/2) \rho_{\rm S\sigma}(E-eV) \rho_{\rm t\sigma}(E) \right. \nonumber \\
& & \hspace{2mm} + \left. \sin^2 (\theta/2) \rho_{\rm S\sigma}(E-eV) \rho_{\rm t\bar{\sigma}}(E) \right\} ,
\end{eqnarray}
with
\begin{widetext}
\begin{eqnarray}
\tilde D(E) & = & \left[ 1 - v^2 g_{\mathrm{SS}, \uparrow \uparrow, 11}(E-eV) 
\left\{ g_{\mathrm{tt}, \uparrow \uparrow, 11}(E) \cos^2(\theta/ 2) +
g_{\mathrm{tt}, \downarrow \downarrow, 11}(E) \sin^2(\theta/ 2) \right\} \right] \nonumber \\
& \times & \left[ 1 - v^2 g_{\mathrm{SS}, \downarrow \downarrow, 11}(E-eV) 
\left\{ g_{\mathrm{tt}, \uparrow \uparrow, 11}(E) \sin^2(\theta/2) +
g_{\mathrm{tt}, \downarrow \downarrow, 11}(E) \cos^2(\theta/2) \right\} \right] \nonumber \\
& - & v^4 \cos^2(\theta/2) \sin^2(\theta/2) g_{\mathrm{SS}, \uparrow \uparrow, 11}(E-eV)
g_{\mathrm{SS}, \downarrow \downarrow, 11}(E-eV) 
\left\lbrace g_{\mathrm{tt}, \uparrow \uparrow, 11}(E) - g_{\mathrm{tt}, \downarrow \downarrow, 11}(E) \right\} .
\end{eqnarray}
\end{widetext}
where the expressions of the different bare Green's functions appearing here can be found in Eq.~(\ref{eq-impurity-dressed}).
Notice that this modified tunneling formula is very similar to the original one, see Eq.~(\ref{eq-I_tunnel}), the only
difference being the presence of the denominator $|\tilde D(E)|^2$. This denominator takes into account the possible
normal reflections in the tunneling between the bound states and renormalizes things to ensure that the transmission
is bounded by 1. In Fig.~\ref{fig-tunnel2} we illustrate that this formula qualitatively captures the crossover mentioned above.
In this figure we show the evolution with the normal state conductance $G_\mathrm{N}$ of the height of the direct 
Shiba-Shiba peak for positive bias, $d^+$, for the set of parameters specified in the caption. The normal state conductance was 
varied by changing the tunnel coupling $v$ and $G_\mathrm{N}$ was computed by evaluating the slope of the current for $eV \gg 2\Delta$. 
As one can see in Fig.~\ref{fig-tunnel2}, see dashed line, Eq.~(\ref{eq-I_tunnel_den}) describes the crossover to a sublinear behavior 
for a sufficiently high normal state conductance, while it reproduces the linear behavior in the deep tunnel regime. For completeness, 
we have also included in Fig.~\ref{fig-tunnel2} the exact result computed with the full formalism of the previous section. 
Notice that the result of Eq.~(\ref{eq-I_tunnel_den}) reproduces the exact results for values of $G_\mathrm{N}$ as high as
$10^{-3}G_0$. This demonstrates that the crossover in this example is all about single-quasiparticle processes and Andreev 
reflections, some of which are actually possible in this voltage range (see next section), play no essential role in the
height of the current peak for the range of $G_\mathrm{N}$ values explored in that figure.

\section{YSR states and multiple Andreev reflections} \label{sec-MARs}

\begin{figure}[t]
\includegraphics[width=0.95\columnwidth,clip]{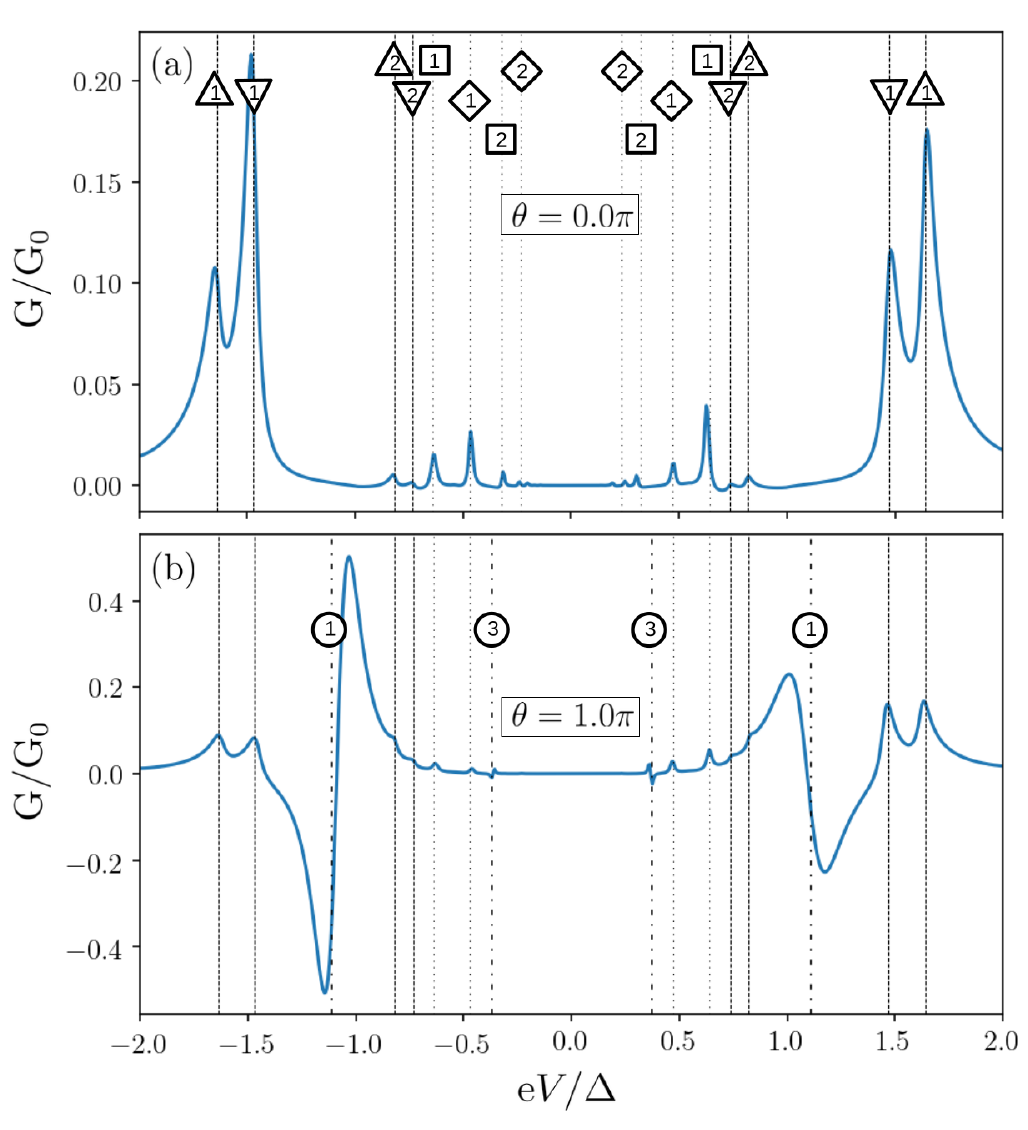}
\caption{Differential conductance $G$ as a function of the bias voltage $V$ for parallel, panel (a), and antiparallel spins, 
panel (b), normalized by the quantum of conductance $G_0 = 2e^2/h$.
The values of the different model parameters are $\Delta_{\rm S} = \Delta_{\rm t} = \Delta$, $\Gamma_{\rm S} = \Gamma_{\rm t} = 
100 \Delta$, $U_{\rm S} = 60 \Delta$, $U_{\rm t} = 0$, $J_{\rm S} = J_{\rm t} = 60 \Delta$, $k_\mathrm{B}T = 0$, $\eta_{\rm S} = 
\eta_{\rm t} = \eta = 0.01\Delta$, and $v = 10\Delta$. With the parameters, the YSR bound states have energies $\pm \varepsilon_\mathrm{S} 
= \pm 0.64\Delta$ and $\pm \varepsilon_\mathrm{t} = \pm 0.48\Delta$, as calculated from Eq.~(\ref{eq-YSR1}). The vertical lines indicate the 
values of several relevant energies. The lines labeled with triangles pointing up and triangles pointing down correspond to processes 
that involve just one YSR state in the impurity $\mathrm{S}$ and $\mathrm{t}$, respectively, and have threshold voltages 
equal to $eV = \pm (\varepsilon_{\rm S} + \Delta)/n$ and $eV = \pm (\varepsilon_{\rm t} + \Delta)/n$ with $n=1,2,\dots$. The processes 
for $n=1$ are single-quasiparticle tunneling and those for $ n \geq 2$ correspond to Andreev reflections of order $n$. The lines 
labeled with squares and diamonds correspond to processes that start (or end) at an YSR state and end (or start) inside the gap region 
due to residual DOS because of finite $\eta$. The threshold voltages are $eV = \pm \varepsilon_{\rm S}/n$ (squares) and 
$eV = \pm \varepsilon_{\rm t}/n$ with $n=1,2,\dots$ (diamonds), depending on whether the YSR state is in the substrate (S) or in the 
tip (t). The lines labeled with circles in panel (b) correspond to processes involving the YSR states of both impurities and occurring 
at voltages $eV = \pm{(\varepsilon_{\rm t} +\varepsilon_{\rm S})} / (2n+1)$ with $n=0,1,2,\dots$ The processes for $n=0$ correspond
to the direct Shiba-Shiba tunneling, while those for $n \geq 1$ correspond to MARs of order $2n+1$. In all cases, the number
inside the symbol indicates the order of the corresponding process in the tunneling probability.}
\label{fig-GV}
\end{figure}

In this section we shall discuss the current-voltage characteristics beyond the tunnel regime with the goal to elucidate the 
different types of MARs that can take place in this system and to provide simple guidelines on how to identify the signatures 
of these processes. In Fig.~\ref{fig-GV} we illustrate the results for the differential conductance, $G = dI/dV$, for parameter 
values similar to those of Fig.~\ref{fig-tunnel2} and for a value of the hopping matrix element $v = 10\Delta$.
Moreover, we shall focus on the case of zero temperature to simplify the discussion. The two panels correspond to the
two limiting cases of parallel spins ($\theta=0$), panel (a), and antiparallel spins ($\theta = \pi$), panel (b). With the parameters
chosen for this figure, the YSR bound states have energies $\pm \varepsilon_\mathrm{S} = \pm 0.64\Delta$ and
$\pm \varepsilon_\mathrm{t} = \pm 0.48\Delta$. For the parallel case of panel (a), we see the appearance of a rich structure, where 
the most pronounced conductance peaks appear at $eV = \pm (\varepsilon_{\rm S} + \Delta) = \pm 1.64\Delta$ and 
$eV = \pm (\varepsilon_{\rm t} + \Delta) = \pm 1.48\Delta$. Obviously, these conductance peaks arise from single-quasiparticle
tunneling connecting the YSR states of the tip and the substrate and the corresponding continuum density of states (DOS) outside 
the gap in the opposite electrode. These are first-order (in $v^2$) tunneling events that give the main contribution to the transport
for parallel spins and low temperatures (they were already present in the example of Fig.~\ref{fig-tunnel1}(e)). Notice that the
height of these peaks is different for positive and negative bias, which is due to the lack of electron-hole symmetry
in this example. In the subgap region ($eV < \Delta$), there is a series of conductance peaks. 
In particular, we observe peaks at $eV = \pm (\varepsilon_{\rm S} + \Delta)/2 = \pm 0.82\Delta$ and 
$eV = \pm (\varepsilon_{\rm t} + \Delta)/2 = \pm 0.74\Delta$. This strongly suggests that these conductance peaks 
are due to second-order ($v^4$) Andreev reflections that involve a YSR state of one of the electrodes and the continuum DOS of the
same lead. These processes also take place in the case of single-impurity junctions and, as it is known, they lead to peaks
that depend on the bias polarity when there is no electron-hole symmetry, see  Ref.~\cite{Villas2020} and references therein. 
Additionally, one can also see several conductance peaks at $eV = \pm \varepsilon_{\rm S}/n$ and 
$eV = \pm \varepsilon_{\rm t}/n$ with $n=1,2$. We attribute these peaks to processes that start or end in a YSR state and end or
start at the residual DOS inside the gap due to the finite broadening parameters ($\eta_{\rm S,t}$). The peaks for
$n=1$ correspond to single-quasiparticle tunneling, while those for $n=2$ correspond to the lowest-order Andreev reflection.
This type of processes was discussed in Ref.~\cite{Villas2020} in the context of single-impurity junctions and we shall not
pay much attention to it in this work. It is also worth remarking that there is no negative differential conductance (NDC) in
this case, i.e., there are no current peaks. Let us also clarify that thermal Shiba-Shiba tunneling discussed in the previous 
section does not show up in Fig.~\ref{fig-GV}(a) because we are assuming zero temperature. 

In the case of antiparallel spins, see Fig.~\ref{fig-GV}(b), the new characteristics, compared to the parallel case, that appear 
in the differential conductance are NDC features at $eV = \pm (\varepsilon_{\rm S} +\varepsilon_{\rm t}) = \pm 1.12\Delta$ and at 
$eV = \pm (\varepsilon_{\rm S} +\varepsilon_{\rm t})/3 = \pm 0.37\Delta$, which correspond to peaks in the current at those voltages. 
The first features are nothing else than the signature of the direct Shiba-Shiba tunneling discussed in the previous section, which are 
due to single-quasiparticle processes between the YSR states in both impurities. The values of the bias at which the second features
appear strongly suggest that they originate from Andreev reflections (of third order in the tunneling probability) that start and end 
in YSR states in a different impurity. As we shall discuss in more detail below, these processes are forbidden in this example for 
$\theta = 0$ because of the full spin polarization of the YSR states, but they are allowed for any $\theta \neq 0$ and its probability 
is maximized for $\theta = \pi$. This type of MAR processes, which we shall refer to as \emph{Shiba-Shiba MARs}, has no analogue in the 
case of single-impurity junctions \cite{Villas2020}. Notice, in particular, that the NDC associated with these processes is a natural 
consequence of their resonant character. Notice also that in this case the features for positive and negative bias are different, which 
again can be traced back to the lack of electron-hole symmetry in this example.

\begin{figure}[t]
\includegraphics[width=0.95\columnwidth,clip]{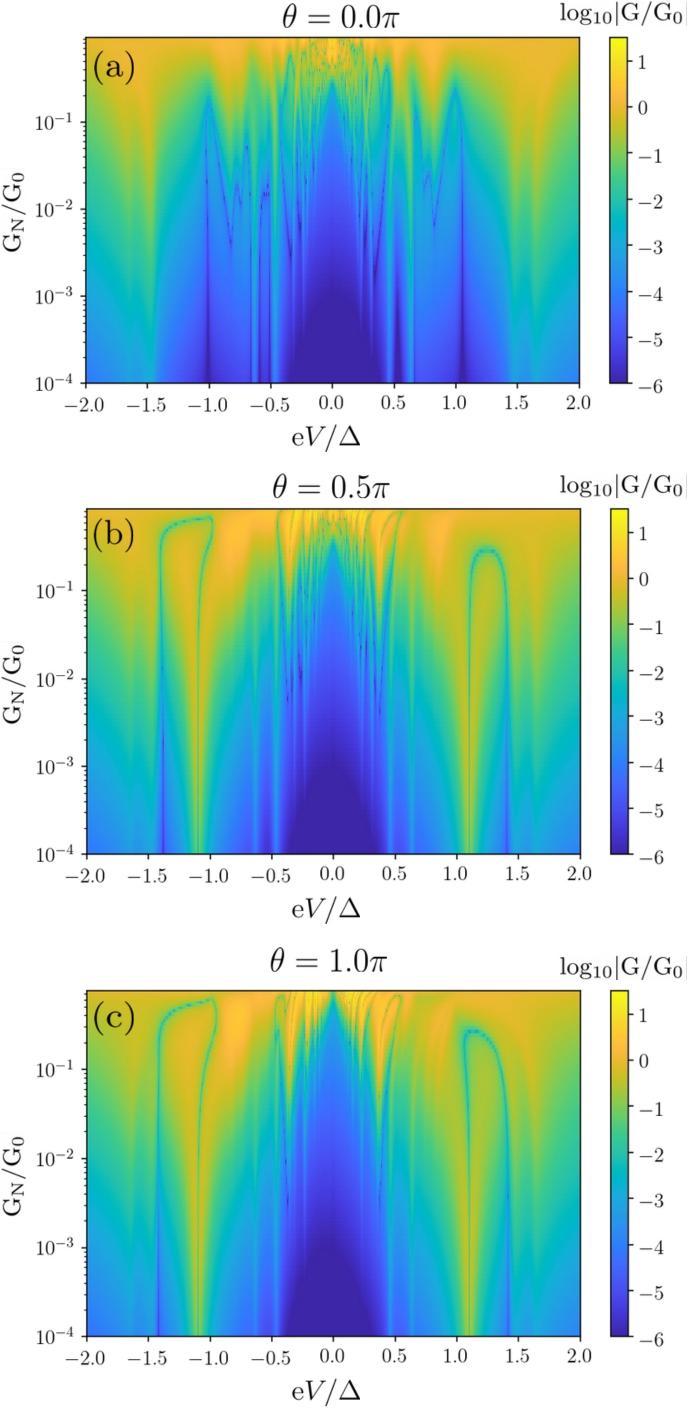}
\caption{Differential conductance as a function of bias voltage and the normal state conductance, $G_\mathrm{N}$, normalized by the
conductance quantum, $G_0$. The different panels correspond to: (a) $\theta = 0$, (b) $\theta = \pi/2$, and (c) $\theta = \pi$. 
The rest of the parameters of the model are $\Delta_{\rm S} = \Delta_{\rm t} = \Delta$, $\Gamma_{\rm S} = \Gamma_{\rm t} = 100 \Delta$, 
$U_{\rm S} = 60 \Delta$, $U_{\rm t} = 0$, $J_{\rm S} = J_{\rm t} = 60 \Delta$, $k_\mathrm{B}T = 0$, and $\eta_{\rm S} = 
\eta_{\rm t} = \eta = 0.01\Delta$.}
\label{fig-GV-map1}
\end{figure}
\begin{figure}[t]
\includegraphics[width=\columnwidth,clip]{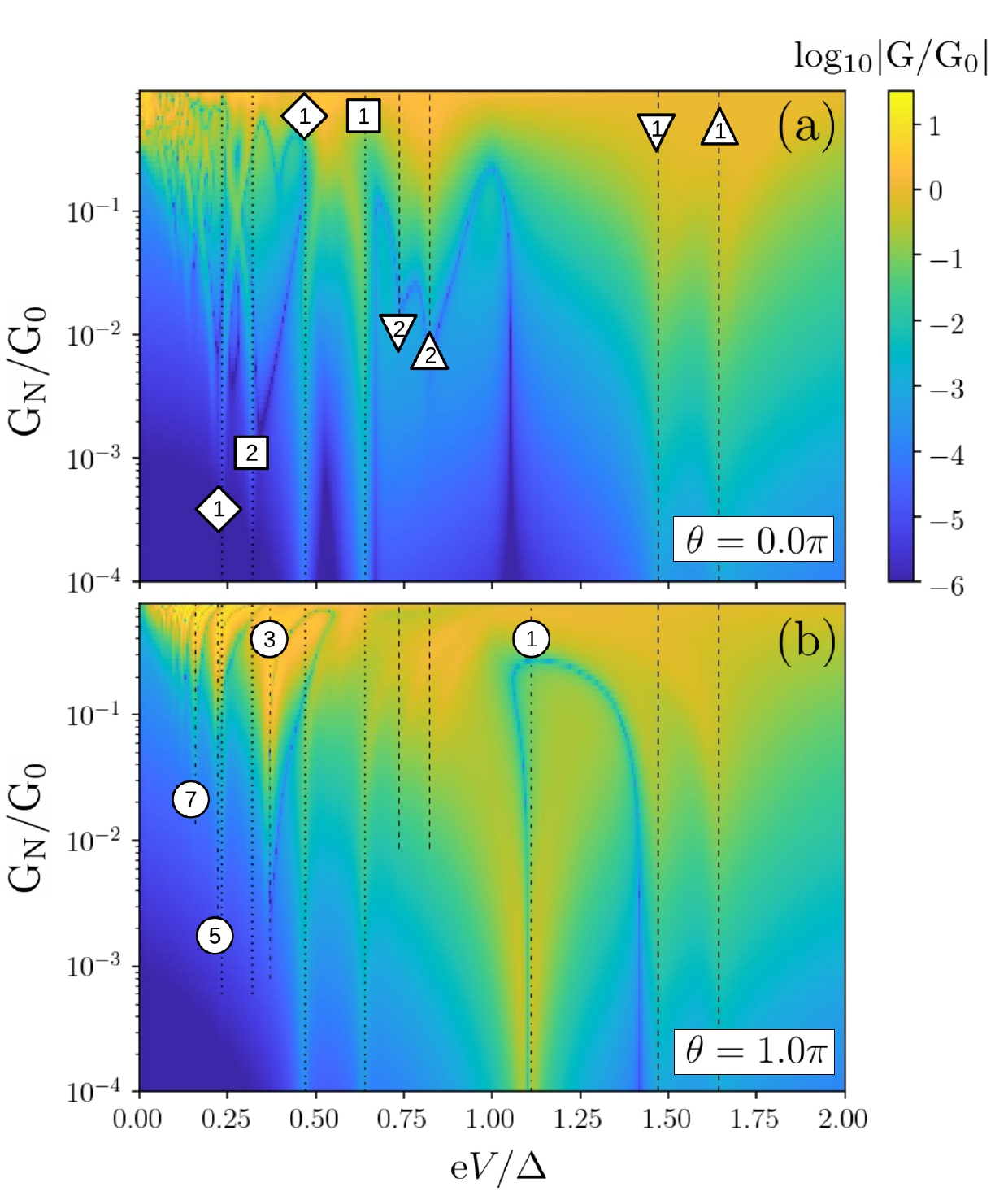}
\caption{The results of panels (a) and (c) of Fig.~\ref{fig-GV-map1} for $\theta = 0$ and $\theta = \pi$, respectively, focusing on 
positive voltages. The vertical lines indicate the values of several relevant energies and the labeling of those lines follows the 
convention of Fig.~\ref{fig-GV}.}.
\label{fig-GV-map2}
\end{figure}

To get further insight into the origin of the subgap features, we present in Fig.~\ref{fig-GV-map1} a systematic study of the
evolution of the differential conductance as a function of the normal state conductance $G_\mathrm{N}$  for the same parameters 
as in Fig.~\ref{fig-GV} (apart from the tunnel coupling), including also the results for an intermediate angle $\theta = \pi/2$. 
Notice that for convenience we are plotting here the absolute value of the conductance in a logarithmic scale. With this choice, 
the NDC appears as a rapid alternation of bright and dark regions. The normal state conductance $G_\mathrm{N}$ was varied in this 
case by changing the hopping $v$ and keeping fixed all the other parameters. In this figure, we can see the evolution of the conductance 
spectra as the junction transmission increases for different values of $\theta$ from the tunnel regime, where only single-quasiparticle
tunneling processes contribute to the transport, to the case of relatively transparent junctions where MAR processes also
contribute giving rise to a very rich subgap structure. To better understand these spectra, we have reproduced the results for
$\theta = 0 $ and $\theta = \pi$ in Fig.~\ref{fig-GV-map2} focusing on positive bias and we have included different vertical lines
indicating the relevant energies discussed in the previous paragraphs. The labeling of these lines follows the convention explained in
the caption of Fig.~\ref{fig-GV}. The most important observation is the appearance for $\theta \neq 0$ of several NDC features 
(corresponding to current peaks) at voltages $eV = \pm (\varepsilon_{\rm S} +\varepsilon_{\rm t})/(2n+1)$ with $n=1,2,\dots$, which 
become more and more prominent as the normal state conductance increases. As explained above, the natural explanation for these features 
is the occurrence of a special type of MARs starting and ending in YSR states of a different impurity. On the other hand, irrespective 
of the value of $\theta$, there is also a series of conductance peaks at $eV = \pm (\varepsilon_{\rm S} + \Delta)/n$ and 
$eV = \pm (\varepsilon_{\rm t} + \Delta)/n$ that can be attributed to MARs that involve a YSR in only one of the impurities.

\begin{figure}[t]
\includegraphics[width=\columnwidth,clip]{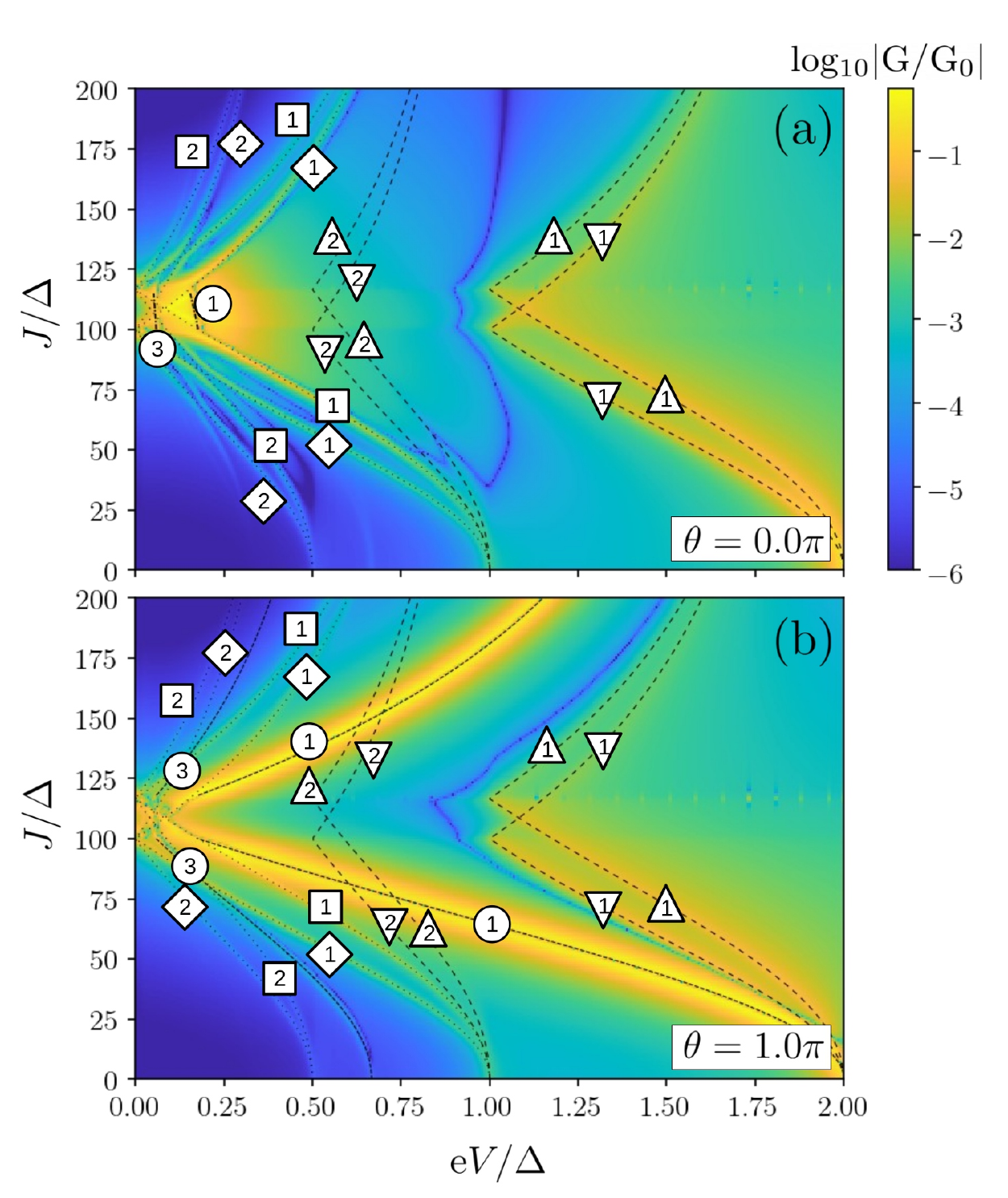}
\caption{Differential conductance as a function of the bias voltage and the exchange energy ($J_{\rm S} = J_{\rm t} = J$) for parallel, 
panel (a), and antiparallel spins, panel (b). The rest of the parameters of the model are $\Delta_{\rm S} = \Delta_{\rm t} = \Delta$, 
$\Gamma_{\rm S} = \Gamma_{\rm t} = 100 \Delta$, $U_{\rm S} = 60 \Delta$, $U_{\rm t} = 0$, $k_\mathrm{B}T = 0$, $\eta_{\rm S} = 
\eta_{\rm t}= 0.01\Delta$, and $v = 5\Delta$. The lines indicate the values of several relevant energies and we follow the labeling 
convention described in Fig.~\ref{fig-GV}.}
\label{fig-J-map1}
\end{figure}
\begin{figure*}[t]
\includegraphics[width=0.9\textwidth,clip]{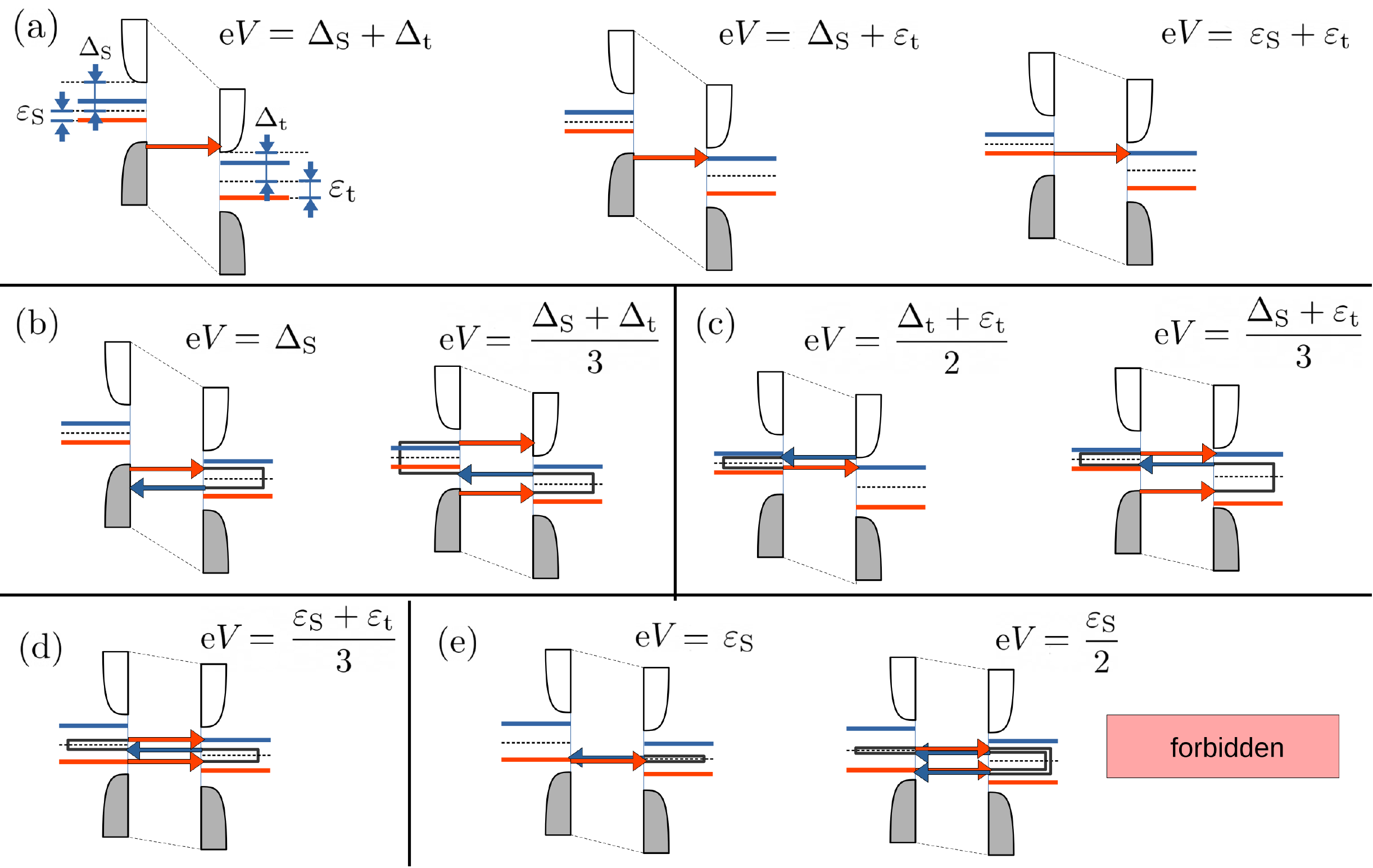}
\caption{Relevant tunneling processes in our two-impurity system. In the different energy diagrams, the left electrode is the 
impurity coupled to the substrate and the right one is the impurity coupled to the tip and their respective density of states are
shifted by the bias voltage. The red lines correspond to electron-like quasiparticles and the blue ones to quasi-holes. In all cases, 
we indicate the threshold voltage at which they start to contribute to the current. (a) Single quasi-particle processes that may involve 
two YSR states (right), one YSR state (center) or none (left). The right one is only allowed when the spins of the two impurities are 
antiparallel. (b) Standard MARs that do not involve any YSR state. (c) MARs that start or end in a YSR state. They give rise to 
conductance peaks at $eV = \pm (\Delta_j + \varepsilon_j)/n$ with $j=\mathrm{t,S}$ when $n>1$ is even or at $eV = \pm 
(\Delta_j + \varepsilon_{\bar j})/n$ when $n>1$ is odd, where $\bar j$ stands for the electrode different from $j$.
(d) MARs that start at a YSR state of one impurity and end in a YSR state of the other impurity. They give rise to the subgap structure 
at $eV = \pm (\varepsilon_{\rm S} + \varepsilon_{\rm t})/n$ where $n > 1$ is odd. (e) MARs that start at a YSR state of one impurity 
and end in a YSR state of the same impurity. They are forbidden due to the full spin polarization of the YSR states.}
\label{fig-processes}
\end{figure*}

To further confirm our interpretation of the origin of the different subgap features, it is convenient to analyze how they shift
when the energy of the YSR states is modified, for instance, by changing the exchange energy. This is what we illustrate in 
Fig.~\ref{fig-J-map1} where we show the evolution of the differential conductance with the exchange energy for the two extreme
cases of $\theta = 0$ and $\theta = \pi$ and for $v = 5 \Delta$. To simplify the analysis we have assumed that both impurities 
have the same value of the exchange energy $J_\mathrm{S} = J_\mathrm{t} = J$, which is changed simultaneously. Notice that we focus 
in this figure on positive voltages simply to make the different features clearly visible. As one can see, there are different 
running lines in these spectra whose dispersion with the exchange energy can be nicely described taken into account the $J$-dependence 
of the energy of the YSR states in both impurities, see Eq.~(\ref{eq-YSR1}). This is illustrated in Fig.~\ref{fig-J-map1} with the 
inclusion of different dotted and dashed lines marking the relevant energies of these features. Thus, for instance, we 
have lines, labeled with circles, that indicate the values of the voltages $eV = \pm (\varepsilon_{\rm S} +\varepsilon_{\rm t})/(2n+1)$ 
with $n=1,2,\dots$, which corresponds to the expected features of the Shiba-Shiba MARs. It may look surprising that some of the
features appearing for $\theta = 0$ have been assigned to these Shiba-Shiba MARs, see circles in panel (a). However, notice that
in those regions, and because of the different values of $U$, the spin of the ground state is different for both impurities and then
the Shiba-Shiba MARs are allowed even for $\theta = 0$. The rest of the features in these spectra that can be attributed to either
the MARs involving a single YSR state in one of the impurities or to the processes involving the residual DOS inside the gap
region. This nicely confirms our interpretations above. Something else that is worth mentioning is the absence, also in the previous 
figures, of the standard subharmonic gap structure at $eV=2\Delta/n$ with $n \in \mathbb{N}$. This structure is due to conventional MARs 
that do not involve YSR states and take place between the continua of states in the leads \cite{Cuevas1996,Villas2020}. In regular 
situations with no impurities, these MARs give rise to the subharmonic gap structure, consisting of conductance peaks at $eV=2\Delta/n$, 
because of the BCS singularities at the gap edges. In our system, those MARs also take place, but the gap edge singularities are not 
present in the DOS of the impurities, which explains the absence of this conventional structure. In an actual experiment, one 
may have additional, non-magnetic channels for tunneling, see e.g.\ Ref.~\cite{Huang2020b}, and then this standard subgap structure 
can coexist with the one we are describing in this work.

After the analysis of the previous results, we are now in position to summarize all the relevant tunneling processes that occur 
in our system, which are schematically shown in Fig.~\ref{fig-processes}. In this figure, the diagrams display the DOS of the substrate
and tip impurities featuring YSR states and we assume a positive bias. Notice, in particular, the absence of gap edge singularities, 
as discussed above. The first class of processes are the single-quasiparticle events shown in panel (a), which dominate the charge 
transport in the tunnel regime. We have three types within this class: (i) tunneling processes between the continua of states in both 
leads (left diagram) with a threshold voltage equal to $|eV| = \Delta_\mathrm{S} + \Delta_\mathrm{t}$, (ii) tunneling processes between 
a YSR state of one impurity and the continuum of states of the other electrode (middle diagram) with a threshold voltage equal to 
$|eV| = \Delta_\mathrm{S,t} + \varepsilon_\mathrm{t,S}$, and the direct Shiba-Shiba tunneling (right diagram) with a resonant voltage 
equal to $eV = \pm (\varepsilon_\mathrm{S} + \varepsilon_\mathrm{t})$. The first type does not produce any abrupt feature (because of 
the absence of gap edge singularities), the second one gives rise to a conductance peak at its threshold voltage, and the third one
is responsible for the direct Shiba-Shiba current peak (with NDC) at its resonant bias. Of course, these processes have their thermal
counterparts at sufficiently high temperature and, in particular, one can have thermally activated tunneling between the YSR at 
$eV = \pm (\varepsilon_\mathrm{S} - \varepsilon_\mathrm{t})$, as we discussed in Sec.~\ref{sec-tunnel-regime}.

The second type of tunneling processes are the conventional MARs shown in Fig.~\ref{fig-processes}(b) that do not involve any YSR
state. As discussed above, these processes usually give rise to a series of conductance peaks at subharmonics of combinations
of the gaps \cite{Ternes2006}, but in our case those features are not visible due to the absence of gap edge singularities. However, 
these MARs can give resonant contributions, where their probability is greatly enhanced, when during the cascade of reflections a
quasiparticle hits the energy of a YSR state in one of the impurities. Thus, for instance, the probability of the second-order Andreev
reflection in Fig.~\ref{fig-processes}(b) is resonantly enhanced when $eV = \pm (\Delta_\mathrm{S} + \varepsilon_\mathrm{t})$. Thus, 
this Andreev reflection competes with the single-quasiparticle process connecting the continuum of states in the substrate impurity and 
the YSR state in the tip impurity, and it eventually dominates the conductance peak height at this bias when the junction transmission is 
sufficiently high. These resonant Andreev reflections take also place in the case of single-impurity junctions where the competition
just mentioned has been discussed in great detail both experimentally and theoretically \cite{Ruby2015,Villas2020}.

A more interesting family of MARs is that described in Fig.~\ref{fig-processes}(c) in which the process starts or ends in a YSR state of 
one of the impurities. Depending on whether the order of the MAR, $n$, is even or odd, one can have two types of threshold voltages
\cite{Holmqvist2014}: (i) $eV = \pm (\Delta_j + \varepsilon_j)/n$ with $j=\mathrm{t,S}$ when $n>1$ is even and (ii) $eV = \pm 
(\Delta_j + \varepsilon_{\bar j})/n$ when $n>1$ is odd, where $\bar j$ stands for the electrode different from $j$. The even processes 
start and end in the same electrode, as in the left diagram in Fig.~\ref{fig-processes}(c), while the odd processes start and end in 
the different electrodes, as in the right diagram in Fig.~\ref{fig-processes}(c). These MARs mediated by a YSR state give rise to 
conductance peaks (with no NDC) at those threshold voltages, as we have illustrated above for the case of a junction with equal 
superconducting gaps. 

\begin{figure}[t]
\includegraphics[width=\columnwidth,clip]{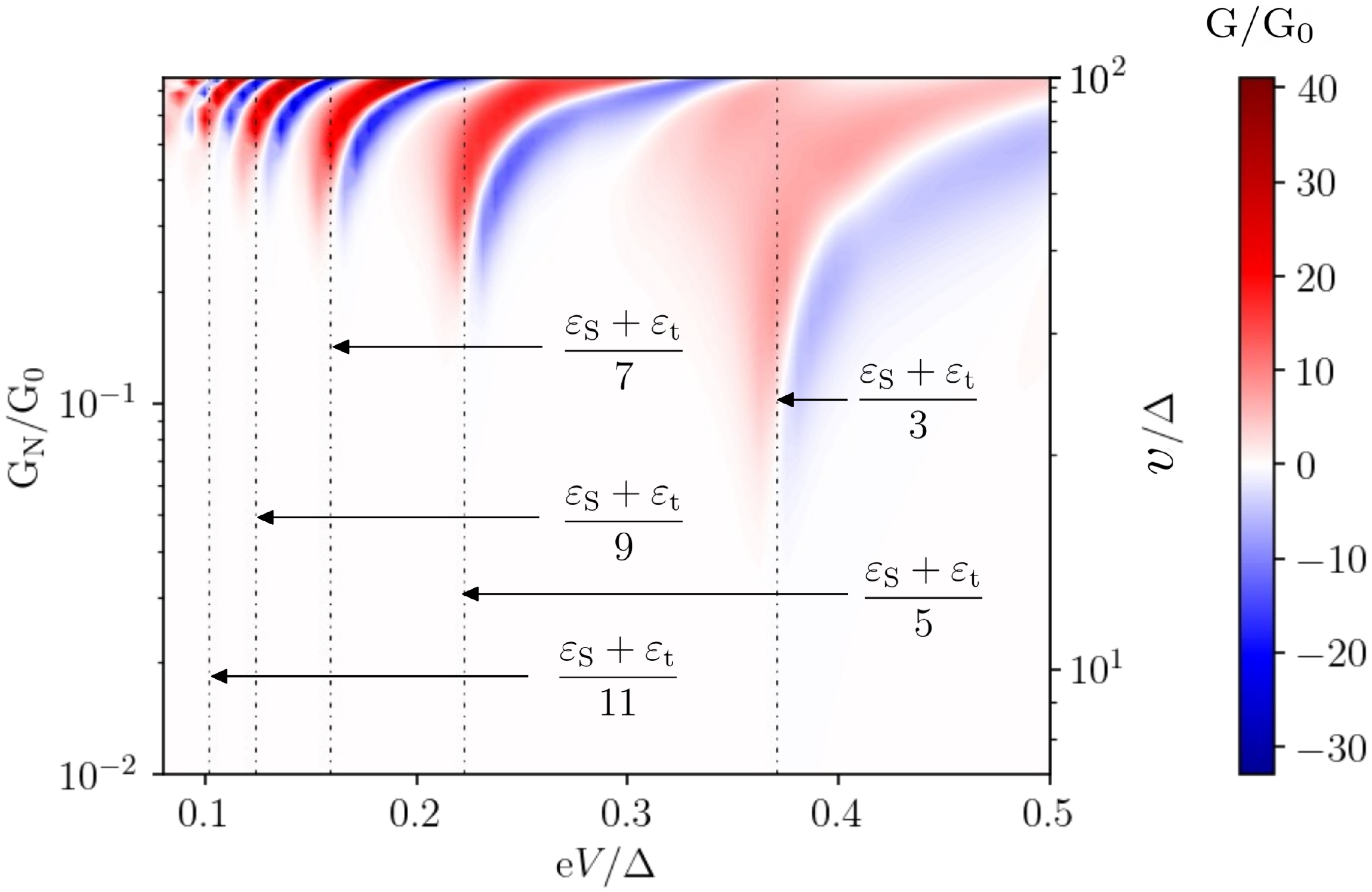}
\caption{Differential conductance in linear scale as a function of the bias voltage and normal state conductance (left 
vertical scale) or hopping matrix element (right vertical scale). The parameters of the model are $\Delta_{\rm S} = 
\Delta_{\rm t} = \Delta$, $\Gamma_{\rm S} = \Gamma_{\rm t} = 100 \Delta$, $U_{\rm S} = 60 \Delta$, $U_{\rm t} = 0$, 
$J_{\rm S} = J_{\rm t} = 60 \Delta$, $k_\mathrm{B}T = 0$, $\eta_\mathrm{S} = \eta_\mathrm{t} = 0.01\Delta$, and $\theta = \pi$. 
The vertical lines indicate the values of several relevant energies corresponding to Shiba-Shiba multiple Andreev reflections. 
The energies of the YSR states are assumed to the ones of the uncoupled impurities given by Eq.~(\ref{eq-YSR1}).}
\label{fig-Shiba-Shiba-MARs}
\end{figure}

Probably the most interesting processes are the MARs that start and end in the YSR states of the different impurities, see
Fig.~\ref{fig-processes}(d). These Shiba-Shiba MARs occur at voltages given by $eV = \pm (\varepsilon_\mathrm{S} + 
\varepsilon_\mathrm{t})/(2n+1)$ with $n=1,2,\dots$ and give rise to current peaks (with NDC) at those voltages. Obviously,
as in the case of the direct Shiba-Shiba tunneling, the width of the current peaks depends on the broadening of the involved
YSR states. To illustrate once more the signature of these peculiar processes, we show in Fig.~\ref{fig-Shiba-Shiba-MARs} 
the differential conductance in linear scale (and no absolute value) for one of the examples that we have discussed above,
but focusing on low bias and relatively high normal state conductance values. In this case, we have used a different
color code for the conductance map to highlight the NDC. As one can see, there is a series of NDC features associated with
these Shiba-Shiba MARs. It is also interesting to notice that those features (corresponding to current peaks) tend to shift 
to higher voltages as the normal transmission of the junction is increased. We attribute this to the fact that for those normal 
state conductance values the electronic coupling between the impurities is strong enough to renormalize the energies of the 
bound states. In other words, those shifts are a signature of the hybridization of the YSR states in the two impurities.
This is an interesting issue that we shall address in detail in a forthcoming paper. 

Finally, we want to mention the MARs shown in Fig.~\ref{fig-processes}(e), which would start and end in a YSR bound state of the 
same impurity. In principle, these processes are energetically allowed and they could give rise to current peaks at $eV = \pm 
\varepsilon_j/n$ ($j=\mathrm{t,S}$) with $n \geq 1$. However, as discussed in Ref.~\cite{Villas2020}, the fact that the YSR states are 
fully polarized makes them forbidden. Such MARs would require a bound state to have a finite DOS of both spin species, which is not 
the case for YSR states.

\section{Conclusions} \label{sec-conclusions}

In summary, motivated by the very recent experimental realization of the tunneling between YSR states, we have presented
in this work a comprehensive theoretical study of the tunneling processes that can take place in a system composed of two 
magnetic impurities coupled to their respective superconducting electrodes. Our analysis is based on the use of a mean-field
Anderson model to describe the magnetic impurities and the Keldysh formalism to compute the current-voltage characteristics.
First, we have shown that our model naturally explains all the basic experimental observations reported so far \cite{Huang2020a},
which concerns the tunnel regime. In this regime, the subgap current exhibits current peaks with very large negative differential 
conductance that are the result of direct and thermally activated tunneling of single quasiparticles between the YSR states in 
both impurities. More importantly, we have predicted that upon increasing the junction transmission, the current can exhibit an 
extremely rich structure in the gap region due to the occurrence of several families of multiple Andreev reflections. Most notably,
we have shown that one can have Andreev reflections connecting the YSR bound states in different impurities and that they give
rise to a series of current peaks at subgap voltages. These processes have no analogue in single-impurity junctions and they
illustrate the new physics that appears when there are superconducting bound states with broken spin symmetry. In principle,
the experimental system of Ref.~\cite{Huang2020a} is ideally suited to test the different predictions put forward in this work.   

\acknowledgments

The authors would like to thank Alfredo Levy Yeyati, Joachim Ankerhold, Ciprian Padurariu, and Bj\"orn Kubala for insightful 
discussions. A.V.\ and J.C.C.\ acknowledge funding from the Spanish Ministry of Economy and Competitiveness (MINECO) (contract 
No.\ FIS2017-84057-P). This work was funded in part by the ERC Consolidator Grant AbsoluteSpin (Grant No.\ 681164) and by the Center 
for Integrated Quantum Science and Technology (IQ$^\textrm{\small ST}$). R.L.K., W.B., and G.R.\ acknowledge support by the DFG 
through SFB 767 and Grant No.\ RA 2810/1. J.C.C.\ also acknowledges support via the Mercator Program of the DFG in the frame of 
the SFB 767.

$^{\dagger}$These authors contributed equally to this work.

\end{document}